\documentclass[]{spie}  

\usepackage{amsmath,amsfonts,amssymb}
\usepackage{graphicx}
\usepackage[table]{xcolor}
\usepackage[colorlinks=true, allcolors=blue]{hyperref}

\title{ANDES, the high-resolution spectrograph for the ELT: design and performance analysis of the YJH spectrograph}

\author[a]{Vinooja Thurairethinam}
\author[a]{David Lee}
\author[a]{James Stevenson}
\author[b]{Étienne Artigau}
\author[b]{Frédérique Baron}
\author[a]{Ciarán Breen}
\author[a]{Bella Boyd}
\author[c]{Denis Brousseau}
\author[d]{David Buscher}
\author[b]{Neil J. Cook}
\author[b]{René Doyon}
\author[d]{Martin Fisher}
\author[a]{Bradley Frank}
\author[a]{Vincent Geers}
\author[a]{Rory Gillespie}
\author[a]{Oscar Gonzalez}
\author[a]{Éamonn J. Harvey}
\author[d]{Wei Li}
\author[d]{Roberto Maiolino}
\author[b]{Lison Malo}
\author[e]{Thomas Marquart}
\author[a]{David Montgomery}
\author[a]{Alan O’Brien}
\author[f]{Ernesto Oliva}
\author[b]{Jonathan St-Antoine}
\author[a]{Jay Stephan}
\author[c]{Simon Thibault}
\author[b]{Philippe Vallée}
\author[a]{Chris Waring}
\author[a]{Sandi Wilson}
\author[a]{Junyi Zhou}
\author[g]{Paolo Di Marcantonio}

\affil[a]{UK Astronomy Technology Centre, Royal Observatory Edinburgh, Blackford Hill, Edinburgh, United Kingdom}
\affil[b]{Université de Montréal, C. P. 6128, Succ. Centre-ville, Montréal, Québec, H3C 3J7, Canada}
\affil[c]{Université Laval, 2375, rue de la Terrasse, Québec City, Québec, G1V 0A6, Canada}
\affil[d]{Cavendish Laboratory, University of Cambridge, JJ Thomson Avenue, Cambridge, CB3 0HE, United Kingdom}
\affil[e]{Uppsala University, Box 516, 75120 Uppsala, Sweden}
\affil[f]{INAF – Osservatorio Astrofisico di Arcetri, Largo E. Fermi 5, I-50125 Firenze, Italy}
\affil[g]{INAF - Osservatorio Astronomico di Trieste, via G. B. Tiepolo 11, 34131 Trieste, Italy}

\authorinfo{Please send any correspondence to V. Thurairethinam, vinooja.thurairethinam@stfc.ac.uk}

\begin{document} 
\maketitle

\begin{abstract}
The ArmazoNes high Dispersion Echelle Spectrograph (ANDES) is a powerful second-generation high-resolution spectroscopic instrument for the Extremely Large Telescope (ELT). The UBV, RIZ, and YJH modules comprise fibre-fed spectrographs of the ANDES baseline design and will offer continuous wavelength coverage of 0.35-1.8 $\mu$m, with the addition of a K-band channel providing coverage up to 2.4 $\mu$m. Coupled with a spectral resolution of $\sim$100,000, ANDES must deliver the required wavelength calibration stability of 1 m/s over 24 hours, with a goal of 0.02 m/s across 10 years. These requirements establish the framework for the infrared module of ANDES, the YJH Spectrograph, leading to what will likely be the largest cryogenic, ultra-stable, high-resolution spectrograph ever built, and will offer the unique ability to observe in both seeing- and diffraction-limited modes interchangeably. We present the current design and performance analysis of the ANDES YJH Spectrograph, outlining the engineering challenges encountered alongside the corresponding strategies adopted to navigate them. In particular, we detail the technology development of the primary dispersing element, an echelle grating mosaic that will span over a metre in length.
\end{abstract}


\section{INTRODUCTION}
\label{sec:intro}  
\begin{figure} [ht]
   \begin{center}
   \begin{tabular}{c} 
   \includegraphics[height=9.4cm]{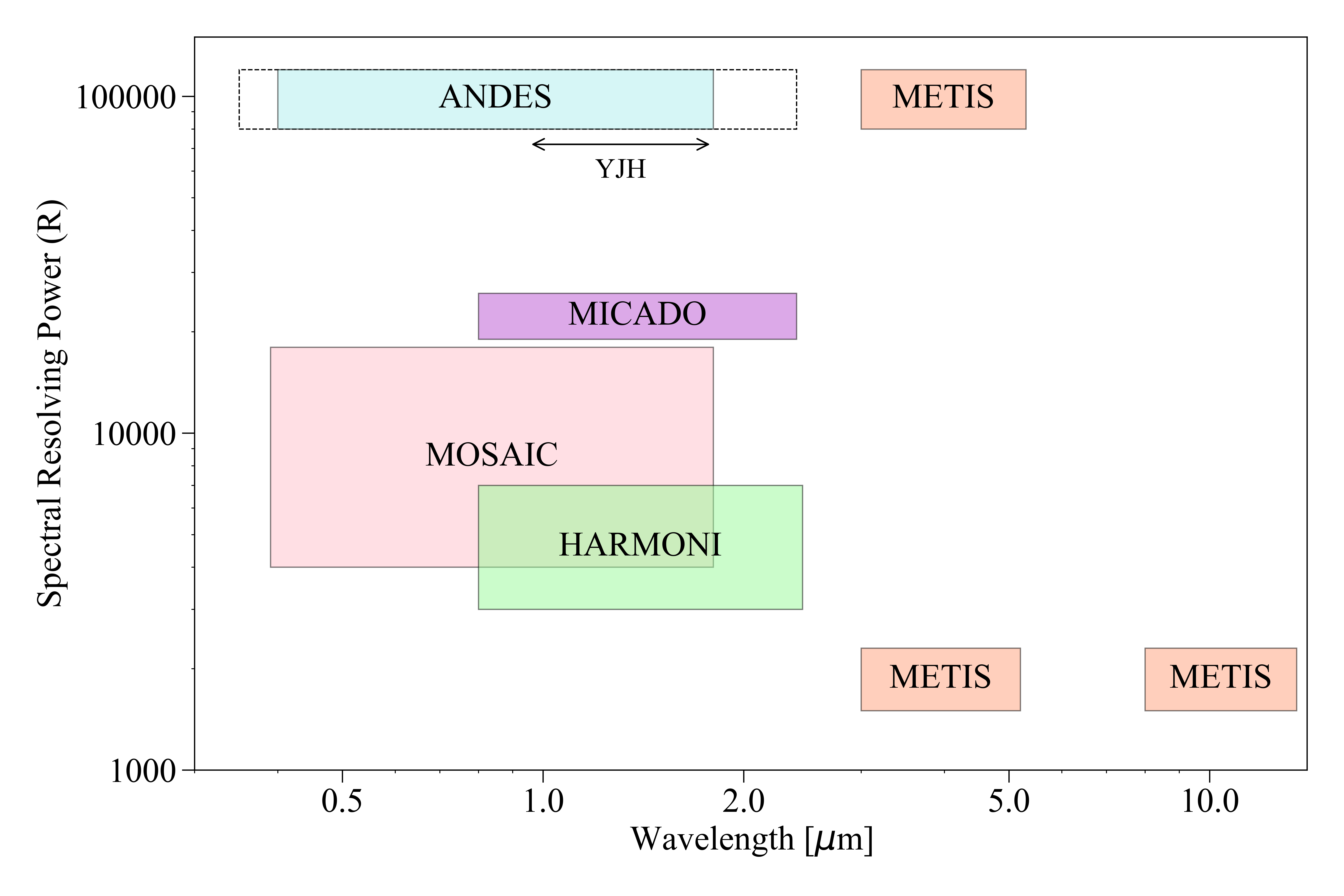}
   \end{tabular}
   \end{center}
   \caption[ELT instruments] 
   { \label{fig:elt_instruments} 
The parameter space of wavelength coverage versus spectral resolution that is occupied by a number of ELT instruments, demonstrating the high-resolution domain and broad waveband that ANDES covers. The wavelength range of the YJH Spectrograph is marked by an arrow and the goal parameter space of ANDES is shown as a dashed box.}
   \end{figure}

The ArmazoNes high Dispersion Echelle Spectrograph (ANDES) is a second-generation, high-resolution visible and near infrared (NIR) spectrograph for the Extremely Large Telescope (ELT), covering wavelengths of 0.4-1.8 $\mu$m at a spectral resolution of R$\sim$100,000 \cite{2024andes_overview}. The construction proposal for the ELT had cited two primary flagship cases as the driver for its build: the detection of biosignatures in Earth-like exoplanets and the direct detection of the acceleration of cosmic expansion \cite{exopl_white_paper}. While the primary science case of ANDES focuses on exoplanet atmospheres, additional science goals include protoplanetary disks, stellar physics and stellar populations, including Population III stars \cite{stars_white_paper}.  Additionally, ANDES aims to probe the epoch of reionisation through observations of the distribution of neutral and ionised hydrogen in the intergalactic medium (IGM) \cite{galaxy_white_paper}. Finally, the spectrograph will offer a look into cosmology and fundamental physics, including probing cosmic expansion as well as searching for potential variations of the fine-structure constant ($\alpha$) and proton-to-electron mass ratio ($\mu$) across various environments and redshifts \cite{cosmology_white_paper}. Such detailed spectroscopic studies warrant a high-resolution spectrograph operating over a wide waveband on a telescope with a large photon-collecting area. Fig. \ref{fig:elt_instruments} illustrates the parameter space of wavelength coverage versus spectral resolution that is occupied by ANDES, in comparison to other ELT instruments. Amongst these instruments, alongside the Integral Field Unit (IFU) mode on the Mid-infrared ELT Imager and Spectrograph (METIS), ANDES dominates the high resolution domain while covering a broad simultaneous waveband from the visible to the NIR. In conjunction with the wide simultaneous wavelength coverage at high spectral resolution and its access to the large collecting area of the ELT, a key requirement of ANDES will be its ultra-stability. With this, ANDES on the ELT will enable a new parameter space to be opened up in this domain, overcoming the current signal-to-noise limitations imposed by telescope size and spectral resolution. 

\begin{figure} [ht]
   \begin{center}
   \begin{tabular}{c} 
   \includegraphics[height=8cm]{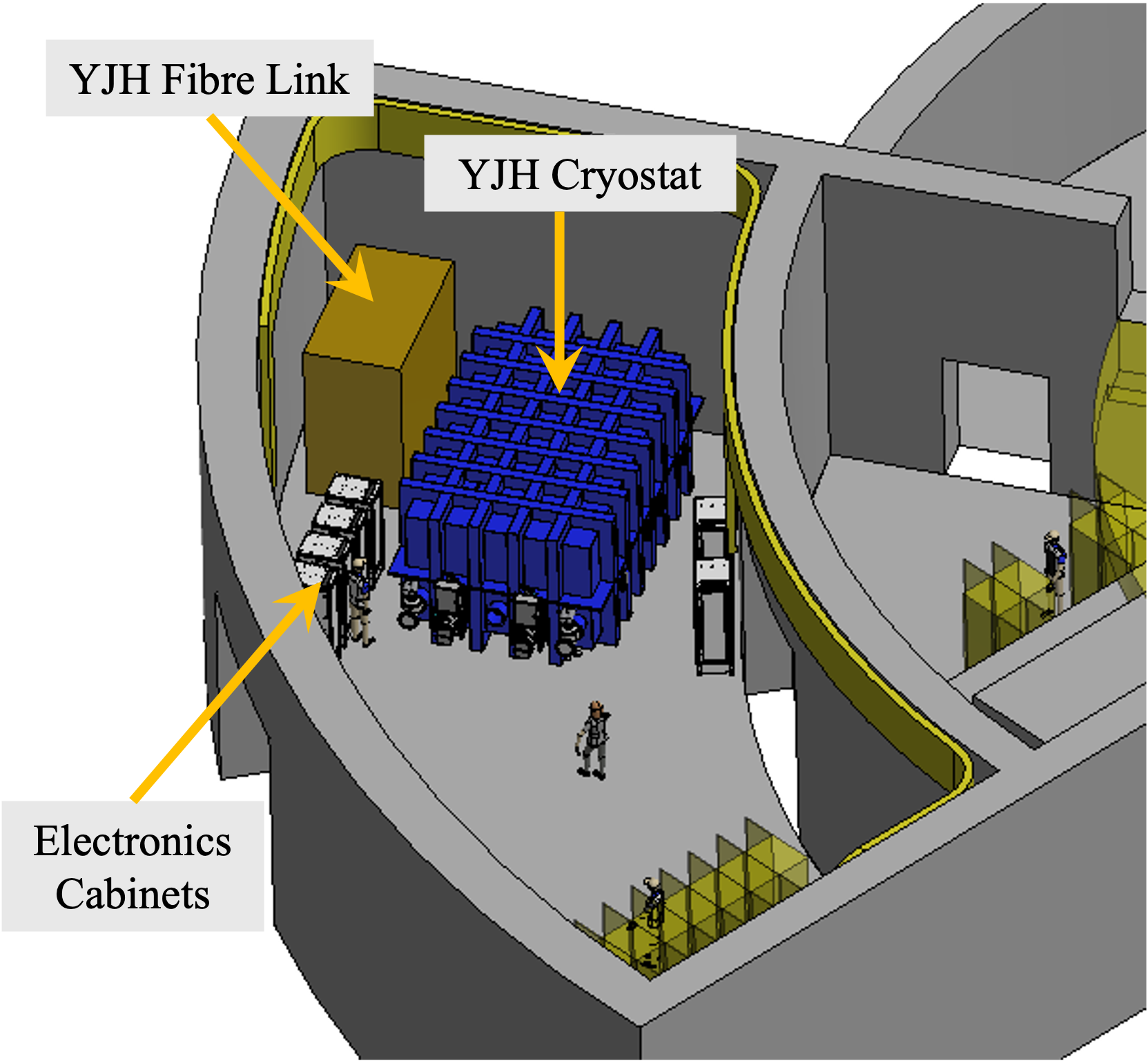}
   \end{tabular}
   \end{center}
   \caption[YJH in Coude] 
   { \label{fig:yjh_coude} 
Current location of the YJH Spectrograph, shown in blue, in the ELT Coudé room, with six electronics cabinets distributed on either side of the volume, as well as the Fibre Link volume for the YS.}
   \end{figure}

The ANDES baseline design is that of a modular fibre-fed echelle spectrograph. This modularity and fibre-feeding of ANDES will allow for modules to be split between the Nasmyth platform and the Coudé room of the ELT. The spectrographs are comprised of the UBV, RIZ, and YJH modules, offering continuous wavelength coverage of 0.35-1.8 $\mu$m, with an additional K-band channel providing coverage up to 2.4 $\mu$m. The YJH spectrograph (YS) will operate at the astronomical Y, J, and H bands from 0.95 to 1.8 $\mu$m and will be located in one of the Coudé rooms that also contains the RIZ module. The YS will be installed in the outer segment of this room alongside the YS electronics cabinets, as shown in Fig. \ref{fig:yjh_coude}. Some of the electronics cabinets, notably the Detector Sub-system cabinet, will need to be installed close to the spectrograph as a result of restrictions on the maximum cable lengths, which have been imposed to avoid an increased impact from noise contributions. Taking into account the telescope infrastructure installed on the ceiling, a clear height of around 6.0 metres is available within the Coudé room. However, the primary restriction on the maximum YS cryostat dimensions is the 4-by-4 metre access door to the room. The relevant parts of the Fibre Link (FL) subsystem will also be mounted next to this cryostat, with the relevant sections of the Calibration Unit (CU) on the far side of the room. 

In addition to transferring light from the ANDES Front End (FE) to a series of fibre slits mounted in the spectrograph, the FL also acts to transmit light from relevant parts of the CU subsystem, providing calibration light to the relevant fibres in the fibre slits. Amongst the baseline spectrometers on ANDES, YS will support the widest range of distinct modes of operation, including a seeing-limited (SL) mode as well as a diffraction-limited, integral field unit (IFU) mode. Light is received through the fibre bundles of the FL from the FE module of ANDES, with the fibres reformatted to form two input slits, where each fibre is fitted with a microlens. Similarly to the other spectrograph modules, the YS will have a SL slit, providing sky, object, and calibration light. Observations performed with this mode will be optimised for throughput and stability without the use of adaptive optics.  The additional diffraction-limited slit is designed to provide light via the IFU in the FE subsystem of ANDES and is assisted by single-conjugate adaptive optics (SCAO), capitalising on the unprecedented angular resolution of the ELT. Following this, regardless of the chosen mode of operation (SL or IFU), the beam travels through a common optical system.

\section{Design Overview}

Upon the selection of an operational mode, a cold slit selector is employed to allow light from the active slit to pass through, whilst simultaneously blocking any stray light from inactive slits. The light from the active slit is formed into a virtual slit, collimated, and dispersed by an echelle grating. Following this, the beam is split into Y, J and H channels using transfer optics and dichroics, where each of the three beams reaches a cross-disperser and camera to form an image of the required spectral orders on a Hawaii H4RG-15 detector. The detectors are controlled and read out by the European Southern Observatory's (ESO) Next Generation Controller II (NGCII) systems using control software running on instrument and detector workstations. The YS control software is used to configure the spectrograph and initiate the readout of data from the detectors. This, in turn, is controlled from the ANDES software, which stores, archives, and reduces the detector data to generate the required science spectral data products.

\subsection{Optical Design}

\begin{figure} [ht]
   \begin{center}
   \begin{tabular}{c} 
   \includegraphics[height=8cm]{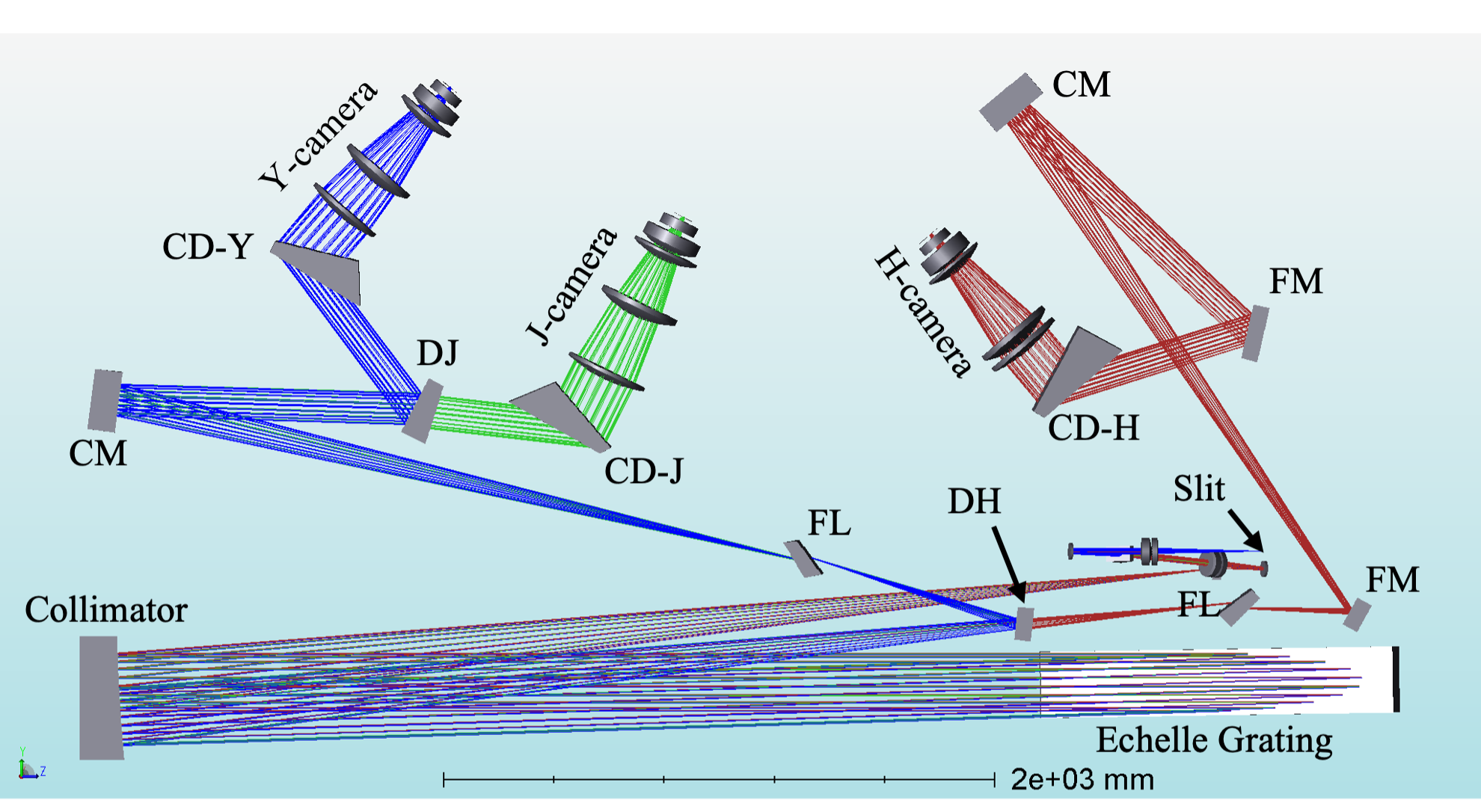}
   \end{tabular}
   \end{center}
   \caption[Optical design] 
   { \label{fig:optical_layout} 
Opto-mechanical model of the V36 ANDES YJH spectrograph. The Y-band beam path is indicated in blue, the J-band in green, and the H-band in red. Some of the larger optics have been labelled as follows: CM – second collimator mirror, CD – cross-disperser, DH – dichroic-H, DJ – dichroic-J, FL – field lens, FM – fold mirror. The small anamorphic slit module optics that directly follow the slit have not been labelled. The optical path for an on-axis field is shown between each optical component.}
   \end{figure}

Each fibre from the FL is fitted with a microlens of 0.6 mm along the slit direction to convert the beam from F/3.5 propagating in the fibre to F/20 required for the spectrograph. An array of 75 microlenses forms a straight, flat slit of length 35.55 mm, with an elliptical chromium mask applied to each microlens to define the required microlens clear aperture of 0.474 mm by 0.436 mm. The baseline slit assembly will consist of two slits, corresponding to the SL and IFU modes. These slits will be stacked above one another, with the appropriate mechanical mounting to connect the full assembly to the rest of the spectrograph. The current version 36 (V36) of the YS is displayed in Fig. \ref{fig:optical_layout}. The beam exits the slit and diverges towards the Anamorphic Slit Module (ASM), which provides anamorphic magnification of the beam and consists of a refractive doublet collimator, a fold mirror, an anamorphic dispersing prism, a refractive doublet camera, a fold mirror, and a wedged plate located at the slit image. The ASM was recently redesigned to move the input slit in the direction of the primary collimator, thereby reducing the overall optical volume and decreasing its length along the z-axis of the YS. This change also provides more space behind the slit assembly to accommodate the minimum fibre bend radius of 300 mm.

The light is then collimated by an off-axis parabolic mirror and reflected towards an R4 echelle grating with 16 lines per mm, located 4000 mm from the collimator at a pupil image. The beam is dispersed by the echelle grating and reflected towards the collimator for a second time. Here, the echelle grating is employed in a slight non-Littrow configuration so that the dispersed beam is reflected towards the collimator at a slight angle to the input beam in the second pass. The beam is then reflected towards the first dichroic filter, Dichroic H, which separates the beam spectrally and spatially into two beams, transmitting wavelengths corresponding to the H-band and reflecting Y and J-band wavelengths. A dispersed image is formed at the YJ or H field lenses, which provide aberration correction. Following the field lenses, the light is recollimated by the secondary collimator mirrors. In the H-band, two flat fold mirrors are needed to redirect the beam. For the Y- and J-bands, the collimated beam is split into the two bands by another dichroic beam splitter. The beam then passes through the cross-dispersers before entering the refractive cameras. Each cross disperser consists of a prism and a transmission grating, which act to separate the otherwise overlapped orders, and thus create a two-dimensional spectrum. The optical designs of the cross-dispersers and cameras were individually optimised for each wavelength channel. 

   \begin{figure} [ht]
   \begin{center}
   \begin{tabular}{c} 
   \includegraphics[height=10cm]{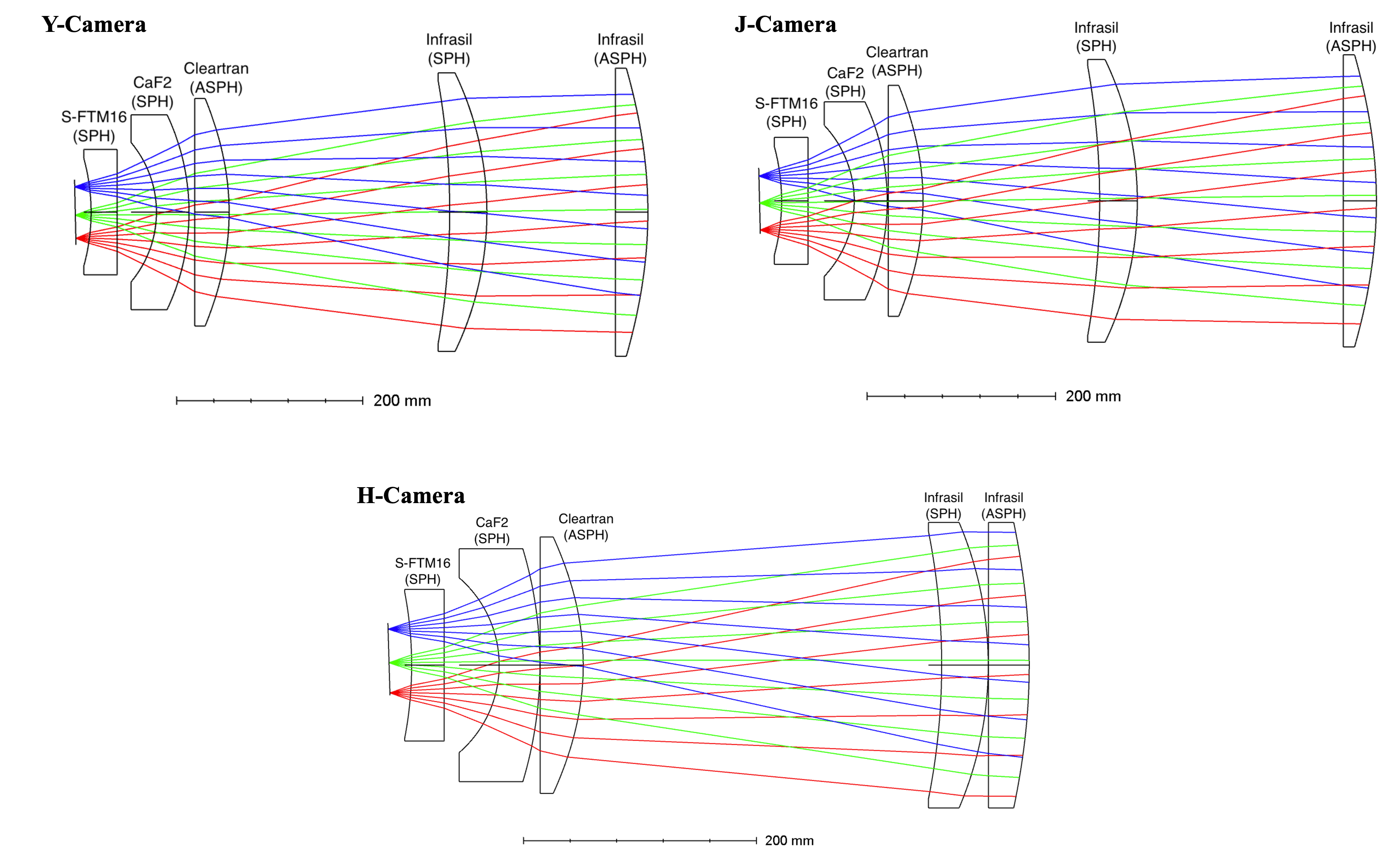}
   \end{tabular}
   \end{center}
   \caption[Camera Y, J, H] 
   { \label{fig:camera YJH} 
Optical ray-trace diagrams of the Y-, J-, and H-band cameras. The material of each lens is specified and classified as either spherical (SPH) or aspherical (ASPH).}
   \end{figure} 

Fig. \ref{fig:camera YJH} shows the optical layout of the Y, J and H band cameras of ANDES. Each of the cameras consists of five lenses, made from the same materials: two Infrasil lenses, one Cleartran, one CaF$_2$, and one OHARA S-FTM16 lens. During the design of these cameras, a custom catalogue of material properties was created to restrict optimisation to a subset of glasses with well-characterised refractive indices and thermo-optical behaviour over the temperature range 100 K to 293 K. The resulting optical materials used in the camera design were based on optimal performance, alongside prior experience with these materials from the Université de Montréal and the Université Laval team from instruments such as the Gemini Planet Imager (GPI), SpectroPolarimètre InfraROUge (SPIRou), and Near Infra-Red Planet Searcher (NIRPS) \cite{GPI_camera, spirou_camera, nirps_camera}. Three of the lenses are spherical, while the front surfaces of the remaining two lenses are eighth-order even aspheric surfaces with conic constants of zero. The initial goal during the optimisation of the camera design was to employ the same five-lenses across the three designs, in order to minimise manufacturing cost, reduce risk, and simplify integration on the optical bench. The current design of the set of five lenses are identical for the Y and J band cameras, while the lenses have marginally different radii and aspherical parameters in the H band. In addition, the air gaps between the lenses vary slightly from band to band. Further information regarding the optomechanical design of a prior version of these camera designs can be found in Ref.~\citenum{2024YJHcamera}.

\subsection{Detector Control}
Each camera includes a Focal Plane Assembly (FPA) featuring a Teledyne H4RG-15 SWIR 2.5 $\mu$m infrared detector, PID-controlled temperature regulation, and opto-mechanical adjustment capabilities (tip-tilt and focus). The detector readout electronics are composed of a cryogenic preamplifier and ESO’s NGCII. Each of the three NGCIIs is controlled by a service application (NGC Core) running on its own Detector Work Station (DWS). Detector control software (DCS) running on an Instrument Workstation (IWS) communicates with each NGC Core application via the network. All software components are fully compatible with the ELT environment. The FPAs are temperature-regulated to better than 1.8 mK over 24 hours to ensure optimal thermal stability. Each detector will undergo optimisation and characterisation prior to integration in Montréal to achieve optimal performance. 

Unlike charge-coupled devices (CCDs), infrared detectors cannot be put in an idling state while accumulating charges. The infrared readout electronics continuously reset or read the charges in individual pixels. Moreover, once the electronics start the reading (or resetting) process, this occurs one pixel at a time and cannot be interrupted until it finishes reading the entire array. It can take up to 3 minutes to read (or reset) the entire pixel array. However, multiple sections of the detector can be read at the same time, and in the case of the YS, the detector will read 64 sections simultaneously. While the H4RG can only record scenes with an integration time equal to the readout time, infrared detectors have the capability to sample charges multiple times without resetting the pixels through non-destructive readout. The YS will employ up-the-ramp (UTR) sampling by starting with a reset and sampling the astronomical scene multiple times at fixed intervals until a complete ramp has been acquired, which occurs once sufficient charge levels have been achieved. With this, the detector will switch to idle mode, consisting of an infinite loop of resets. Contrary to CCDs, the integration time for infrared detectors is defined by a combination of the number of ramps, reads, and resets instead of an absolute integration time.

\subsection{Mechanical Design}

   \begin{figure} [ht]
   \begin{center}
   \begin{tabular}{c} 
   \includegraphics[height=7.5cm]{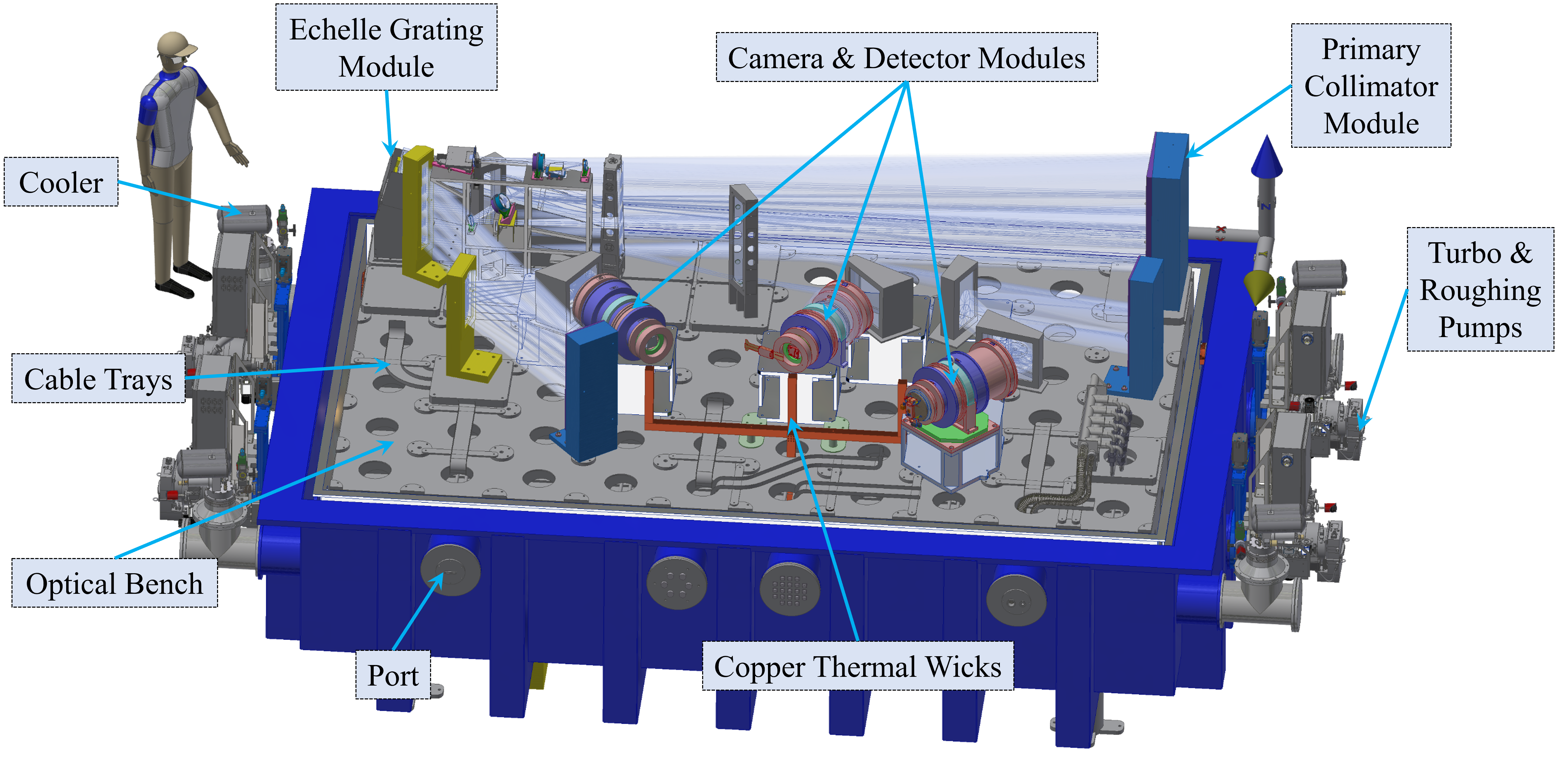}
   \end{tabular}
   \end{center}
   \caption[Mech assembly] 
   { \label{fig:mech_assmbly_cryostat} 
YJH spectrograph vacuum vessel with its upper segment removed. Fibres and cabling are not shown here.}
   \end{figure} 

The general mechanical design has been formulated based on prior and current experience of a number of heritage instruments built at the UK Astronomy Technology Centre — namely the Submillimetre Common-User Bolometer Array 2 (SCUBA2), the Multi-Object Optical and Near-infrared Spectrograph (MOONS), and the High Angular Resolution Monolithic Optical and Near-infrared Integral field spectrograph (HARMONI) \cite{scuba2_cryostat, moons_cryostat,harmoni_cryostat}, the design of which remains under development at the time of writing. The YJH vacuum vessel, housed in a thermal enclosure, will be in the shape of a cuboid made of Aluminium alloy 5083 and supported on four legs. The vacuum vessel, with external dimensions of 5.9 $\times$ 3.8 $\times$ 2.7 m, splits horizontally near the plane of the internal optical bench, such that the upper segment may be removed for internal access, as portrayed in Fig. \ref{fig:mech_assmbly_cryostat}. The lower fixed component of the vessel provides the external interfaces to the support system, vacuum infrastructure, cooling infrastructure, cabling and fibre feedthroughs, as well as internal thermal isolating supports. The optical bench and a single radiation shield are supported off the lower part of the vacuum vessel by three thermally isolating flexures to provide an isostatic support. The cuboid radiation shield composed of Aluminium alloy 6082-T6 is employed and provides a light-tight enclosure around the optical bench with a similar horizontal split at the level of the vacuum vessel. The exterior of this shield will feature a multi-layer insulation (MLI) blanket with the aim of reducing heat loads and thermal gradients on the radiation shield.


The vacuum vessel, required to provide thermal insulation for the cold structure and to provide a clean environment for the detector and optics, is expected to be at a temperature of 300 K within the Coudé room. However, the cryogenic system must maintain the temperature of the bench at 100 K and the radiation shield at 130 K, to ensure the photon noise remains insignificant, whilst simultaneously maintaining the temperature of the detectors at a lower temperature of 80 K. The YS intends to use the Cryomech 810 pulse tube cooler recommended by ESO and take advantage of the two stage cooling that it offers. It has been estimated that a minimum of three such coolers will be necessary to achieve the required temperatures and cooling power. ANDES YJH will employ four pulse tube coolers to cool the radiation shield and detector to yield improved cooling uniformity and symmetry. The resulting maximum loads on the Cryomech 810 pulse tube cooler first and second stages have been deemed acceptable at 90 W and 2.5 W at temperatures of 75 K and 10 K, respectively. Additionally, a liquid Nitrogen pre-cool/warm-up system is incorporated to facilitate the cool-down and warm-up of the instrument. This system utilises a derivative of the ESO standard cooling pads to provide uniform cooling of the optical bench and radiation shield, and is based on comparable systems developed for the Very Large Telescope (VLT) - MOONS and ELT - HARMONI instruments, see e.g. Ref. \citenum{moons_cooling}. A number of temperature sensors will be employed to permit the monitoring of the various temperatures, the exact number and locations of which remain to be determined following further analysis.

\section{Design Considerations}



\subsection{FL-YS Interface}

   \begin{figure} [ht]
   \begin{center}
   \begin{tabular}{c} 
   \includegraphics[height=6cm]{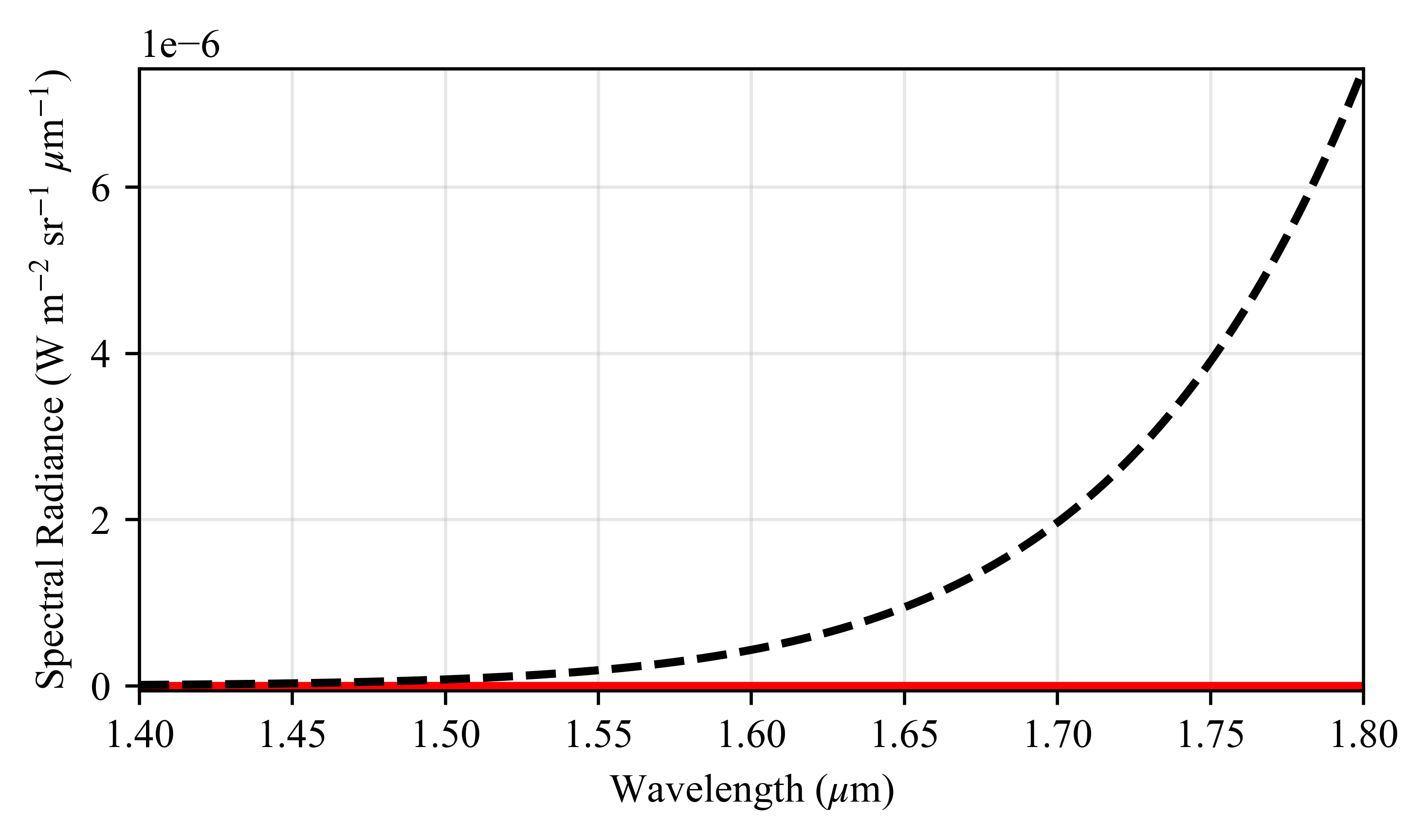}
   \end{tabular}
   \end{center}
   \caption[Blackbody slit selector] 
   { \label{fig:blackbody_slit_selector} 
Example blackbody spectra with and without a cold slit selector. The dashed black line shows an example thermal emission within the H-band with no cold slit selector at a temperature of 291 K. The red line shows the reduced thermal emission with a cold slit selector at a temperature of 130K.}
   \end{figure} 


The FL-to-YJH interface defines the input to the spectrograph, and is therefore a key contributor to the YS performance. This interface covers everything from the incoming optical fibres, through to the slit assembly within the cryostat. As the baseline, the slit assembly comprises an SL slit and an IFU slit. Here, a cold slit selector mechanism allows for the blanking of any given slit to shield the thermal emissions from the FE structure passing through the non-operational slit. This mechanism thus minimises stray light from external sources within the YS, shielding the spectrograph from thermal emission that would otherwise be detectable, particularly in the H-band. An example blackbody spectrum, in the H-band, for a source emitting at 291 K (18 $^\circ$C) is shown in Fig. \ref{fig:blackbody_slit_selector}. The blackbody spectrum shows increasing thermal emission at wavelengths above 1600 nm that would be detectable by the H-band camera, with a spectral radiance of  7.420$\times 10^{-6}$ W/m$^2$/sr/$\mu$m for a blackbody source at 291 K, at a wavelength of 1.8 $\mu$m. If the cold slit selector is cooled to 130 K, its spectral radiance will fall to 1.249$\times 10^{-14}$ W/m$^2$/sr/$\mu$m, thereby providing a factor of $>10^{14}$ reduction in thermal background. Currently, the slit selector is intended to be operated at the same temperature as the optical bench (100 K) but there may be space to relax this  specification as the design matures.

In addition to the selection of a single slit, the cold slit selector will also support a dark position, as well as an open position fitted with a neutral density (ND) filter to facilitate the observation of bright sources while preventing the detectors from saturation. Tests performed on the MOONS detectors demonstrated that the default readout speed of the Hawaii H4RG detector is 200 kHz, using 64 outputs in buffered mode \cite{moons_detectors}. In this configuration, the minimum readout time for a 4096 $\times$ 4096 pixel detector is 1.31 seconds. With Correlated Double Sampling (CDS), two readouts are required, and so, despite the interval between two readouts of a given pixel being only 1.31 seconds, the total minimum CDS readout time is 2.62 seconds. The requirement for minimum exposure time on the YS is 3 seconds, with a goal of 1 second. This means that the requirement can be met directly with this performance, but the goal cannot. The addition of the ND filter mitigates this constraint and allows for the observation of these bright sources without saturating the detector.

   \begin{figure} [ht]
   \begin{center}
   \begin{tabular}{c} 
   \includegraphics[height=5.5cm]{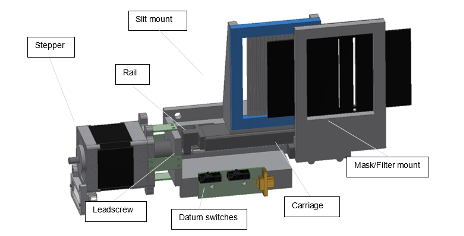}
   \end{tabular}
   \end{center}
   \caption[Slit selector] 
   { \label{fig:slit_selector} 
Mechanical layout of the slit selector mechanism assembly.}
   \end{figure} 
   
A mechanical layout of the slit mechanism is shown in Fig. \ref{fig:slit_selector}, comprised of a fixed mount and a linear moving mask. The mask can block the light from the slit, allow light to pass through a given slit aperture, and has additional positions to hold an ND filter.

\subsection{The Echelle Grating}

The V36 YJH spectrograph optical design employs an echelle grating as the primary disperser at an angle of incidence of 76 degrees, also referred to as an R4 echelle, with a small non-Littrow tilt angle of 0.9 degrees. The grating has a ruling of 16 lines per mm, but it should be noted that this specification only describes the warm dimension as per the catalogue, and will differ marginally when cooled to operating temperature. The grating orders are shown in Table \ref{tab:orders}. 

\begin{table}[ht]
\caption{Diffraction orders covered in each waveband of the YS.}
\label{tab:orders}
\begin{center}       
\begin{tabular}{ccc}
\hline
\hline
\textbf{Band} & \textbf{Diffraction Orders} & \textbf{Number of Orders} \\
\hline
\hline
Y & 109 -- 127 & 19 \\
J & 90 -- 108  & 19 \\
H & 68 -- 83   & 16 \\
\hline
\end{tabular}
\end{center}
\end{table}

   \begin{figure} [ht]
   \begin{center}
   \begin{tabular}{c} 
   \includegraphics[height=4.8cm]{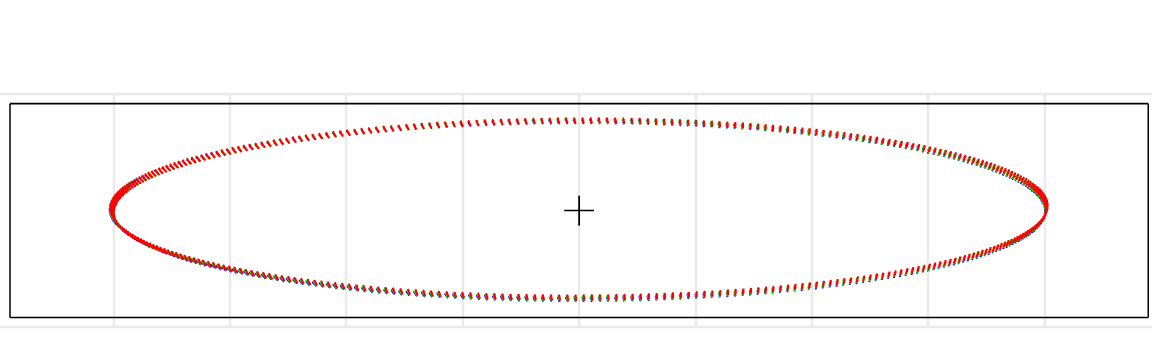}
   \end{tabular}
   \end{center}
   \caption[Echelle Footprint] 
   { \label{fig:echelle_footprint} 
Beam footprint diagram at the echelle grating, for the H-band V36 design. The size of the grating shown is approximately 250 mm by 1130 mm with a beam footprint of 213 mm by 1096 mm.}
   \end{figure} 

The beam footprint on the echelle grating is approximately 213 mm (spatial) by 1096 mm (spectral), for all field points and three slits, as shown in Fig. \ref{fig:echelle_footprint}. Due to the limitations of current manufacturing and alignment processes, this means that the grating cannot be fabricated as a single section and must instead be constructed as a grating mosaic, as has been done for the Echelle SPectrograph for Rocky Exoplanets and Stable Spectroscopic Observations (ESPRESSO) \cite{espresso_echelle_mosaic}. However, it is important to note that the YS grating mosaic differs from ESPRESSO in that it will be operating cryogenically and oriented on its back. The exact number of mosaic segments required for this will depend on the manufacturing capabilities of the chosen supplier. There is expected to be a small gap between each sub-grating in the mosaic, estimated to be 10 mm, and this will cause a small throughput loss. Any stray reflections from the inter-grating gap are likely to be specular and may thus necessitate additional baffling. For this, alternative approaches include using a stepped arrangement of the gratings to avoid vignetting and remove the need for baffles \cite{espresso_echelle_mosaic}. The viability of the grating for supply and manufacture has been a core point of consideration in the design of the YS. 

\begin{table}[ht]
\setlength{\tabcolsep}{3pt}
\caption{Examples of echelle gratings used in other spectrographs: High Accuracy Radial velocity Planet Searcher (HARPS) \cite{harps_grating}, Echelle SPectrograph for Rocky Exoplanets and Stable Spectroscopic Observations (ESPRESSO) \cite{espresso_echelle_mosaic}, Habitable-zone Planet Finder (HFP) \cite{hpf_grating}, WINERED - high-blazed echelle grating (HBG) \cite{winered_grating}, Near InfraRed Planet Searcher (NIRPS) \cite{nirps_grating, nirps_grating_substrate,bouchy2025nirps}, GIANO - Telescopio Nazionale Galileo (TNG) \cite{giano_grating}, SpectroPolarimètre InfraROUge (SPIRou) \cite{spirou_grating}, Near InfraRed SPECtrograph (NIRSPec) - Keck II \cite{nirspec_grating}.}
\label{tab:echelle_gratings_list}
\centering
\small
\begin{tabular}{lcccccccc}
\hline
\hline
\textbf{Instrument} & \textbf{$\lambda$ [$\mu$m] }& \textbf{R} & \textbf{Size [mm]} &
\textbf{Mosaic} & \textbf{Type} & \textbf{lines/mm} & \textbf{Substrate} & \textbf{Cryogenic?} \\
\hline
\hline
HARPS     & 0.378--0.691 & 120k & 214$\times$840$\times$125 & Monolithic & R4 & 31.6 & Zerodur & No \\
ESPRESSO  & 0.380--0.788 & 70k/140k/190k & 1220$\times$204 & 3$\times$1 & R4 & 31.6 & Zerodur & No \\
HPF       & 0.81--1.28   & 53k & 200$\times$800 & Monolithic & R2 & 31.6 & Zerodur & Yes (200 K) \\
WINERED   & 0.96--1.35   & 70k & 200$\times$60 & 2$\times$1 & R5.3 & 11.06 & ULE & No \\
NIRPS     & 0.97--1.81   & 75k/88k & 90$\times$320 & No & R4 & 13.3 & Si & Yes \\
GIANO  & 0.9--2.5     & 50k & 135$\times$365$\times$45 & No & R2 & 23.2 & Fused SiO$_2$ & Yes \\ 
SPIRou    & 0.95--2.35   & 70k  & 154$\times$306$\times$50 & No & R2 & 23.2 & Zerodur & Yes \\
NIRSPEC   & 0.95--5.1    & 20k & 142$\times$320$\times$50 & No & R2 & 23.2 & Al & Yes \\
\hline
\rowcolor{gray!20}
ANDES YJH  & 0.95--1.8 & 100k & 250$\times$1330 & Yes & R4 & 16.0 & Zerodur/Invar & Yes \\
\hline
\end{tabular}
\end{table}

Table \ref{tab:echelle_gratings_list} summarises a list of gratings employed in other instruments, selected based on resolution, wavelength coverage, dimensions or ruling, to illustrate the complexity of the YS echelle grating requirements. Across the reviewed spectrographs, several of the instruments use 31.6 lines per mm R4 echelle gratings. However, as the YS operates at longer wavelengths further into the infrared, the increased dispersion at these wavelengths makes this specific line spacing solution non-viable. The intended 16 lines per mm R4 design presents an equivalent line spacing when optimised for the near-infrared wavelengths. In addition, most of the examples included in Table \ref{tab:echelle_gratings_list} employ gratings that are of much smaller size than that of the YS. Only ESPRESSO has a grating area of comparable size to that of the YS; however, it uses a 31.6 lines per mm grating and is non-cryogenic. Amongst the cryogenic instruments, the majority are non-mosaiced and have a much smaller grating. As such, procurement of the 16 lines per mm R4 echelle grating is considered a high risk to the ANDES project, and the YS team is undertaking a range of de-risking activities, including communicating closely with suppliers. As part of this, an R2 solution based on commercial masters was considered, and a trade study had been carried out. However, this option would have necessitated a two-dimensional mosaic and would have also prevented the spectrograph from complying with its scientific requirements, including the required spectral resolution of 100,000. Ultimately, the R2 option was not deemed to provide an advantage over the R4 solution with regards to de-risking.
As such, the current baseline remains an R4 design, specifically based on an Invar substrate referenced as IC-DX \cite{IC-DX}.

   \begin{figure} [ht]
   \begin{center}
   \begin{tabular}{c} 
   \includegraphics[height=2.95cm]{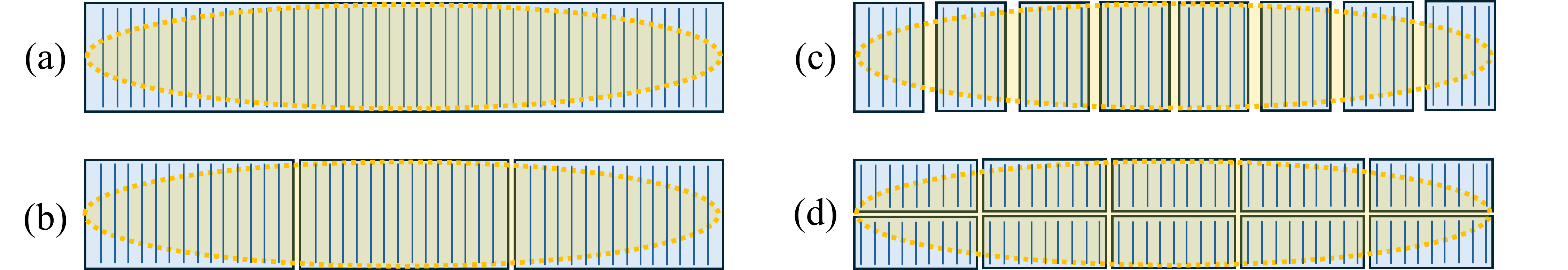}
   \end{tabular}
   \end{center}
   \caption[Mosaic Configuration] 
   { \label{fig:mosaic_config} 
Illustrations of the different grating mosaic configurations. The direction of the grooves is indicated using vertical lines and the beam footprint is represented by an orange ellipse.}
   \end{figure} 

The exact configuration of the mosaic is dependent on the supplier and the grating type selected. Fig. \ref{fig:mosaic_config} shows illustrations of the different configurations under consideration. The first of these, Fig. \ref{fig:mosaic_config}(a) is a single monolithic grating. This has merely been included for completeness and is not a viable option since no manufacturer could make a single grating of this size. A single substrate solution could only be possible by the fabrication of a monolithic mosaic (as used by e.g. HARPS). This consists of a repeated replication process whereby a single master is replicated multiple times onto a single substrate. This process has not been considered, since it would require more than two replications on the substrate. Co-aligning two replicated gratings is a challenging process, and the alignment of possibly one or two additional gratings is unlikely to be a feasible option. An alternative configuration, pictured in Fig. \ref{fig:mosaic_config}(b) and (c), consists of one-dimensional mosaics. This arrangement would be preferable, as the various segments only require alignment in one dimension along the length of the grating. Amongst options (b) and (c), (b) is the preferred solution due to having fewer segments. While (c) may be required depending on the supplier's manufacturing capabilities, this variety would have a lower efficiency than the (b) configuration due to the increased number of segments, resulting in an increase in non-grating material within the beam footprint. The final option shown in Fig. \ref{fig:mosaic_config}(d) is a two-dimensional mosaic. There are, however, many additional concerns that come with the addition of a second dimension to the mosaic. The overall alignment becomes more complex due to the additional alignment required in the second axis, which means that the alignment tolerances and, therefore, stability tolerances, become tighter, particularly across the width of the grating. Moreover, the throughput is further reduced compared to a one-dimensional mosaic due to additional gaps between the mosaic sections. Each mosaic section would require alignment features to be machined into it, and this increased number of sections would also result in increased costs and mechanical complexity. This configuration was only considered for the now-rejected R2 solution.

\subsubsection{Effect of Grating Segment Misalignment on the Spectrograph Line Spread Function}

   \begin{figure} [ht]
   \begin{center}
   \begin{tabular}{c} 
   \includegraphics[height=8cm]{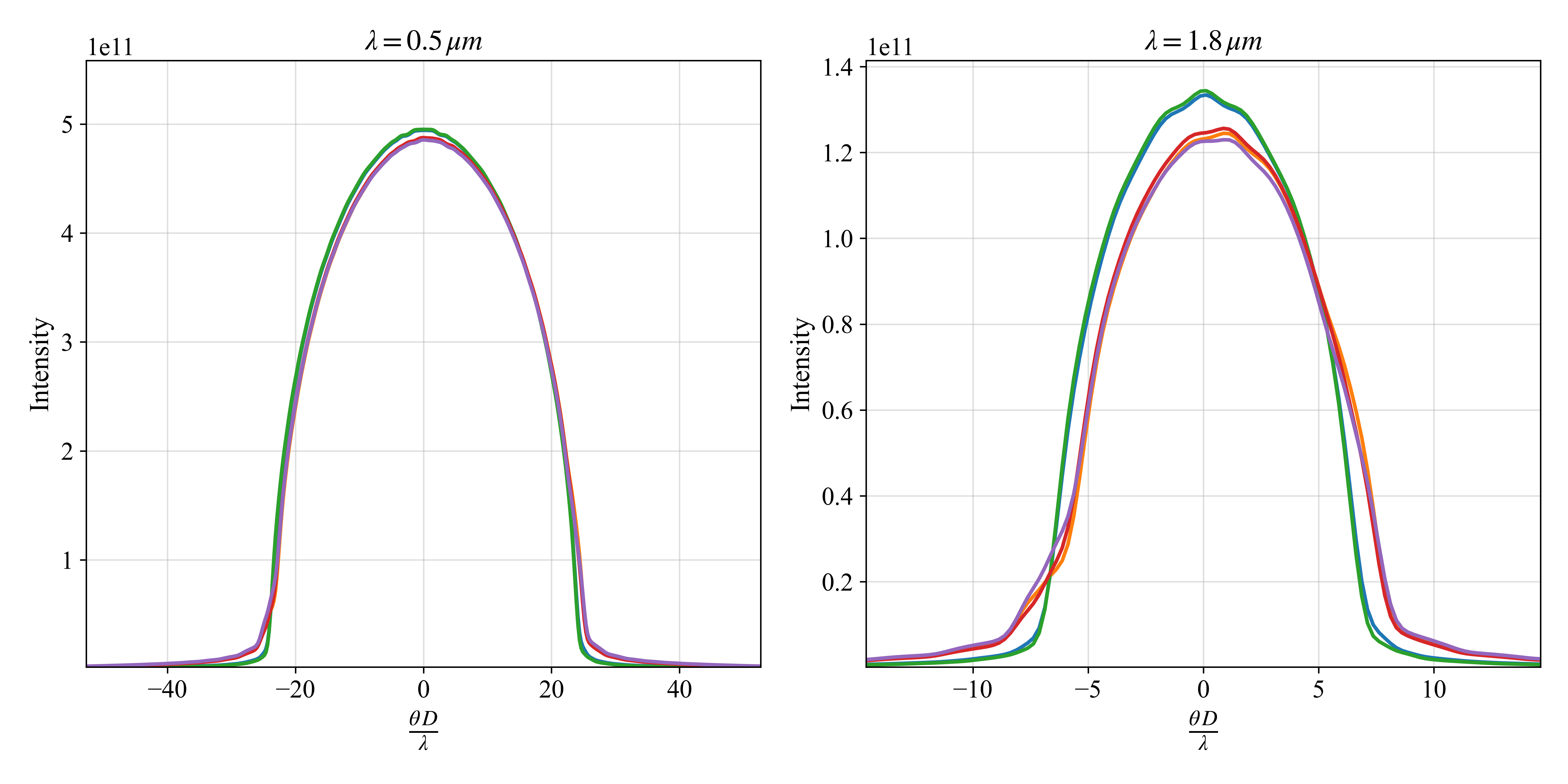}
   \end{tabular}
   \end{center}
   \caption[Grating Misalignment] 
   { \label{fig:grating_misalignment} 
Simulated spectral LSF for a 3 x 1 grating mosaic with random piston offsets assigned to each grating segment at a wavelength of 0.5 $\mu$m (left) and 1.8 $\mu$m (right). Here, $D$ is the diameter of the telescope and $\theta$ is the angular distance from the optical axis.}
   \end{figure} 
   
A simulation investigating the impact of misalignments in a mosaiced/segmented grating on the spectrograph line spread function (LSF) was undertaken at the University of Cambridge. For this, the individual grating segments of a 3 $\times$ 1 mosaic are shifted in piston, in the direction of the illuminating beam, such that the misalignments are defined here as translation errors normal to the blaze direction. While such shifts change the phase uniformly across each of the segments since they remain aligned in tip and tilt, the piston misalignment causes the phases to vary between segments. The simulation assumes a circular-core multi-mode fibre with a size such that a 0.76 arcsecond field on a $D=39$ metre telescope is sampled by a 6 $\times$ 6 array of fibres. At the shortest, visible wavelength of 0.5 $\mu$m of the YS, the diameter of the fibre is 48 $\times$ $\lambda/D$, while the fibre diameter is only 13 $\times$ $\lambda/D$ at the longest NIR wavelength of the H-band of 1.8 $\mu$m.

Fig. \ref{fig:grating_misalignment} illustrates that there is a limited effect of piston misalignment on the LSF at visible wavelengths. At infrared wavelengths, however, the effect of piston phase shifts is noticeable, with clear distortion and spreading in the wings of the LSF. Nonetheless, whether this distortion of the LSF is sufficiently large to cause a significant effect on the current science cases remains yet to be determined and further evaluation of the impact on any potential future science cases is merited.

\subsection{Echelle Grating Mount}

The technology development required to realise the cryogenic echelle mosaic mounting sub-system is currently underway at the University of Cambridge. At the time of writing, there are two designs that address the needs of different substrate types, where the Invar design is considered the baseline. It should be noted, however, that these designs were produced prior to the recent V36 optical design and are expected to be reviewed and further revised in the near future.

\begin{figure}[htbp]
    \centering
    \includegraphics[width=0.47\textwidth]{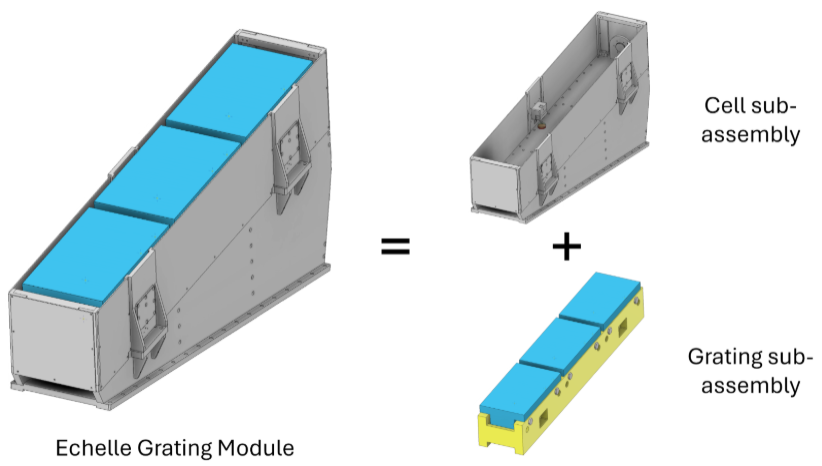}
    \hfill
    \includegraphics[width=0.47\textwidth]{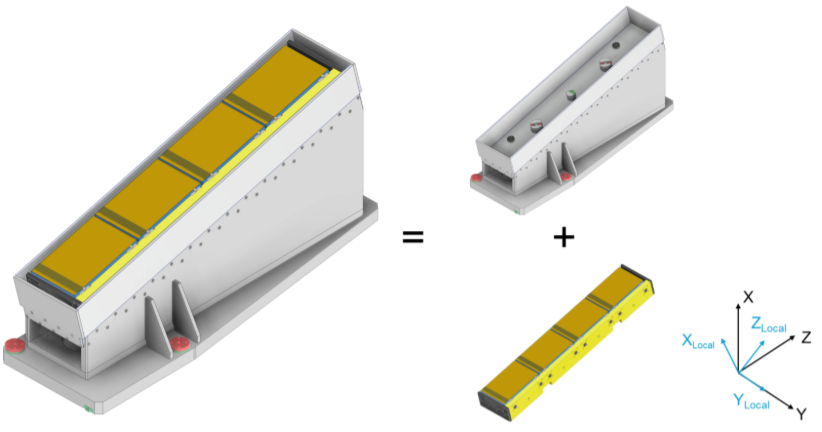}
   \caption[Grating Mount] 
    {\label{fig:grating_mount}
The composition of the grating module and the local coordinate system for two designs currently being explored: the Zerodur version (left) and the Invar IC-DX version (right).}
\end{figure}

The complete grating module comprises a grating sub-assembly and a cell sub-assembly, as shown in Fig. \ref{fig:grating_mount}. By using a single type of material that demonstrates suitable thermal performance for the major components and key alignment parts, the grating mosaics are expected to maintain dimensional stability to meet the alignment requirements at both room temperature and cryogenic temperatures. The grating sub-assembly is assembled and pre-aligned at ambient temperature and is mounted in the aluminium cell sub-assembly using kinematic supports. The entire module can then be bolted to the optical bench of the instrument. The general concept of both solutions of the echelle assembly is very similar. For the Zerodur version, the grating base, shown in yellow in Fig. \ref{fig:grating_mount}, has a “H” geometry aimed at achieving high bending stiffness with low mass. This design was also necessary due to space constraints imposed by nearby optics in pre-V36 designs. The alternative Invar IC-DX grating will be formed as a 4$\times$1 mosaic based on the manufacturer's capabilities, in contrast to the previously considered 3$\times$1 Zerodur-based design. Additionally, the grating base has an “L” geometry, to further conserve space for optical components on the bench surrounding the grating. Analysis is ongoing for the Invar design, particularly looking at the impact of the increased mass in comparison to the Zerodur design. However, IC-DX has demonstrated exceptional thermal performance \cite{IC-DX} and  is easily light-weighted. Moreover, it will likely be much easier to step the gratings since the inter-module hap can be reduced to $\sim$4 mm and all adjustment features can be constructed of the same material.

\section{Simulated Performance}

\subsection{Optical Throughput and Image Quality}

  \begin{figure} [ht]
   \begin{center}
   \begin{tabular}{c} 
   \includegraphics[height=8cm]{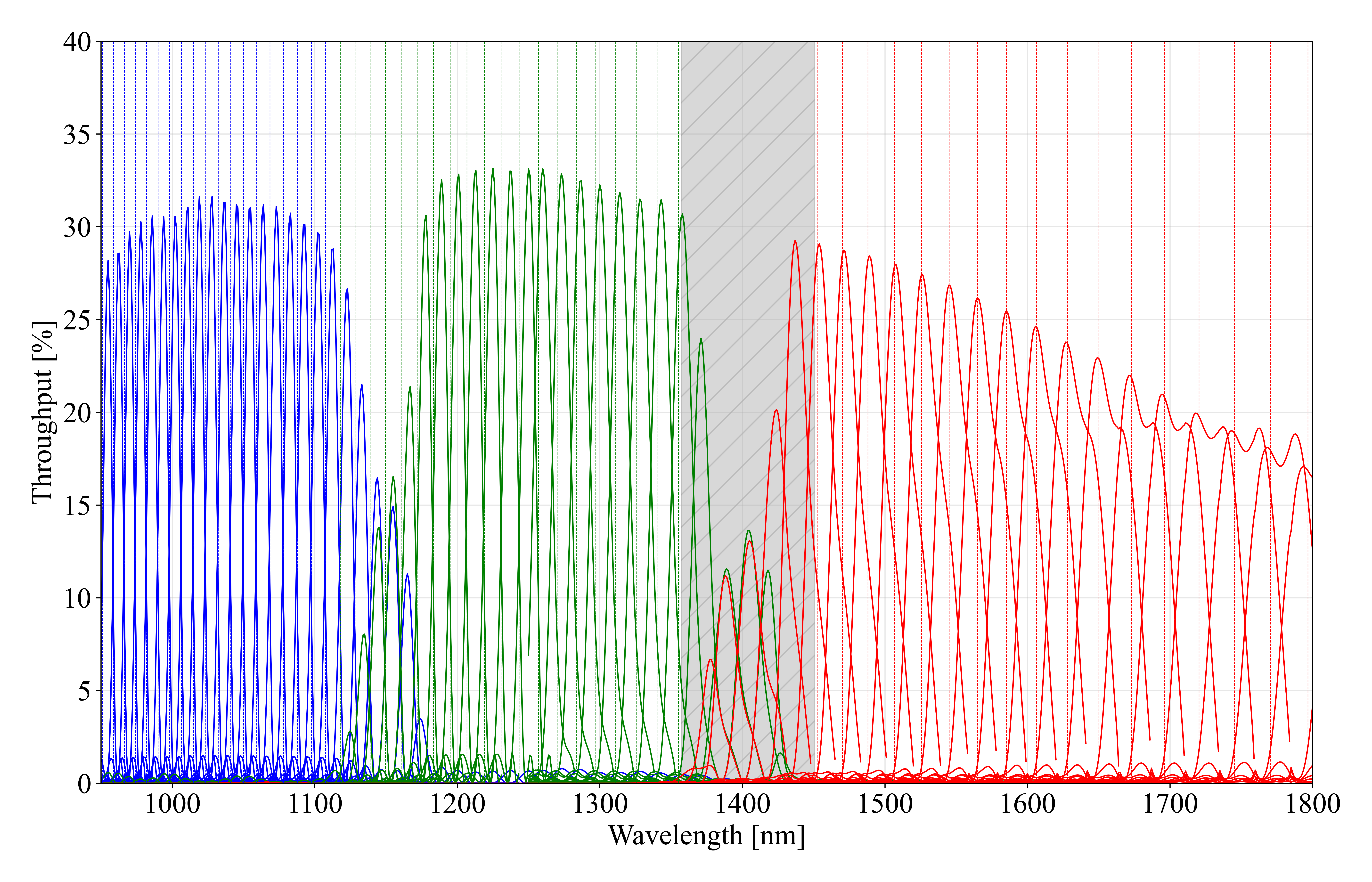}
   \end{tabular}
   \end{center}
   \caption[Throughput] 
   { \label{fig:throughput} 
Total YS throughput as a function of wavelength. The different bands of Y, J and H are shown in blue, green and red, respectively. The vertical dashed lines indicate the extent of wavelengths of each order on a $\pm$30 x $\pm$30 mm detector (allowing for a 0.72 mm margin around the edge of the detector). The grey hatched area marks the region where no throughput requirement is necessary.}
   \end{figure} 

A preliminary throughput estimate for each waveband has been calculated, making top-level assumptions about the reflectivity and transmission of typical gratings, lenses, and mirrors. The dichroic beamsplitters are modelled using existing data. Dichroic J, with a cut-on wavelength of around 1150 nm, has been modelled using commercially available data for similar off-the-shelf components. Dichroic H, with a cut-on wavelength of around 1400 nm, has been estimated using data from the H Dichroic of the MOONS instrument provided by the supplier. Similarly, the remaining optics are modelled using available off-the-shelf components and coatings, as a conservative estimate of their optical performance. The efficiency of the echelle grating was largely simulated using PCGrate, a specialised software used for the modelling of diffraction gratings. The nominal diffraction efficiency of the echelle grating design has thus been simulated to have a groove spacing of 16 lines/mm and a collimated beam incident at 76 degrees. The model further assumes a groove facet angle of 90 degrees and a gold coating. The efficiency of the cross-dispersers has been modelled in Zemax OpticStudio using the Kogelnik approximation for a volume phase holographic surface, and the wavelength-dependent quantum efficiency (QE) of the detectors was taken from Ref \citenum{Detector_report}. Lastly, a “throughput margin” of 5\% has been applied as a contingency allowance for optical components that may be non-compliant with their requirements. With this, Figure \ref{fig:throughput} shows the total YS throughput as a function of wavelength. A more detailed throughput estimate will be calculated following interactions with coating suppliers and will also include losses due to surface contamination and vignetting, including losses caused by the gaps between the echelle grating mosaic segments. 

\begin{figure}[htbp]
    \centering
    \includegraphics[width=0.32\textwidth]{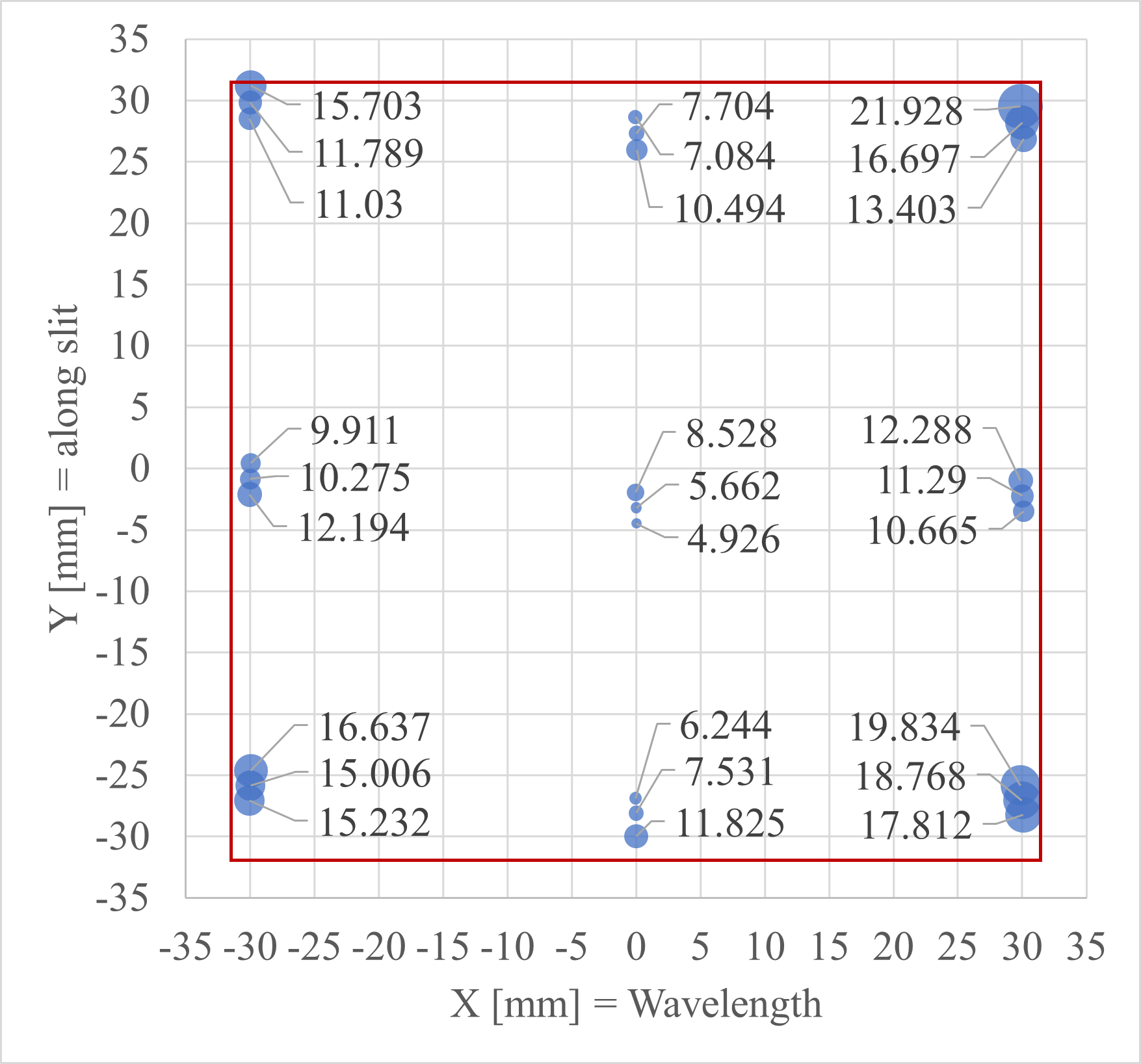}
    \hfill
    \includegraphics[width=0.32\textwidth]{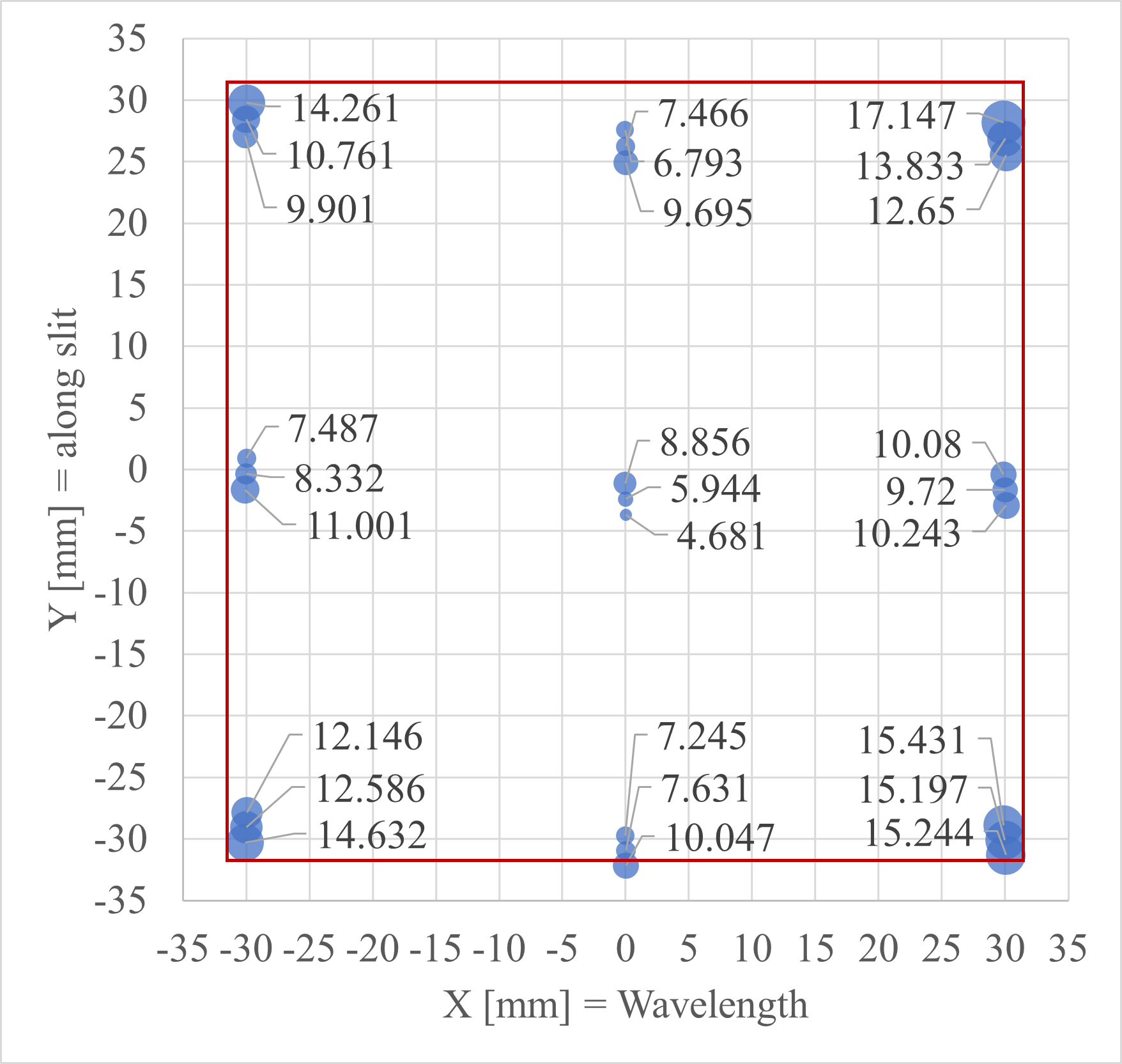}
    \hfill
    \includegraphics[width=0.32\textwidth]{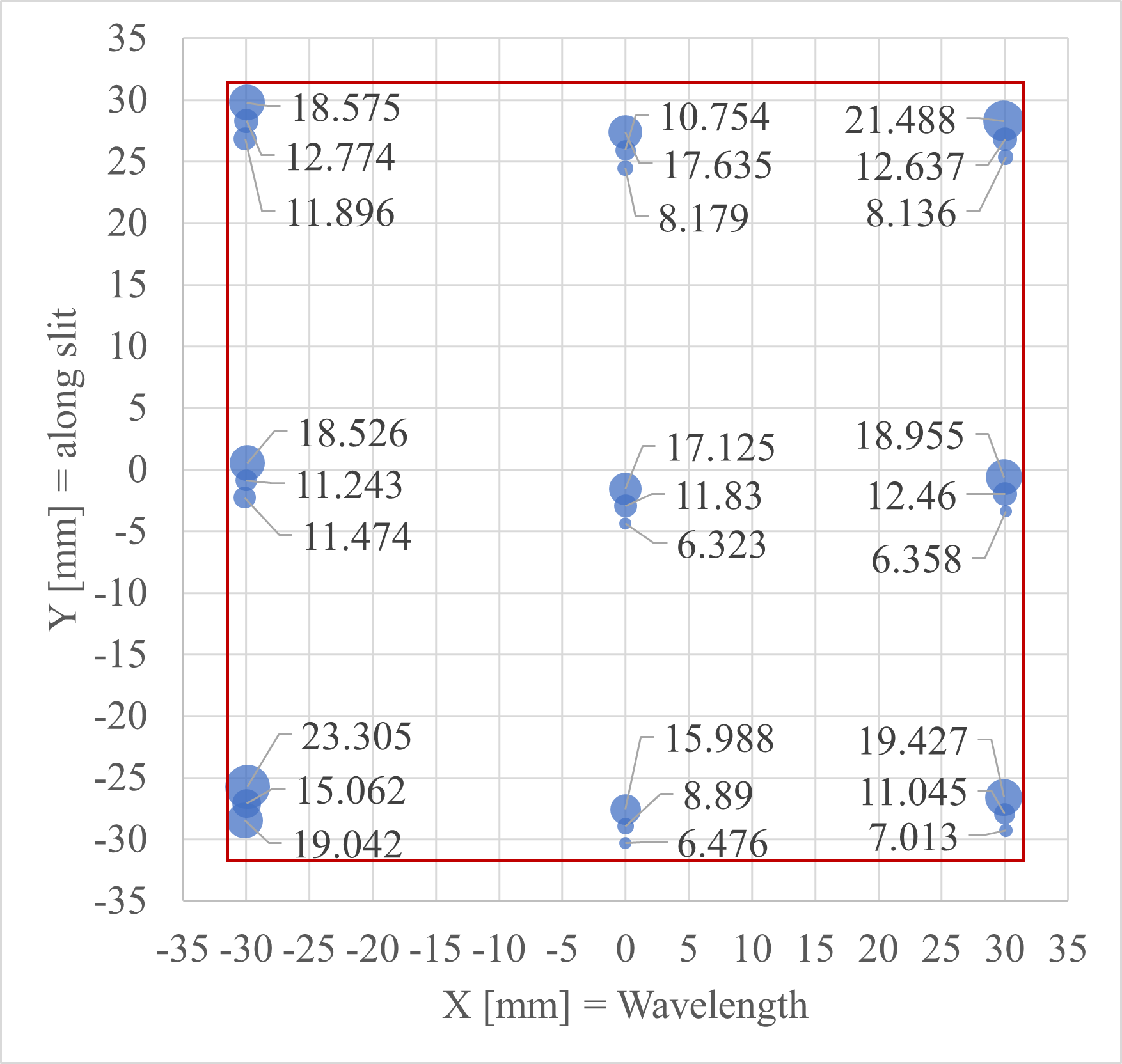}
   \caption[Image Quality] 
    {\label{fig:image_quality}
Bubble plot of RMS spot radius in microns versus image position on the detector for the various wavebands of the V36 YS design. The width of each circle represents the RMS spot radius at that location and wavelength increases from bottom to top, and right to left. The red square indicates the approximate size of the detector. The labels of the bubbles indicate the RMS spot radius in micrometres, with bubble width scaling is set to 15.}
\end{figure}

The image quality was found to vary across the position on the detector, as well as along the slit for a given position. A bubble plot, where the size of the circles represents the nominal RMS spot radius, is shown in Fig. \ref{fig:image_quality} for Y, J and H-band. The change in image quality over the field of view of the detector can be seen, generally with better image quality in the centre of the detector and some degradation in image quality in the corners, along with a change in image quality along the slit. 

\subsection{Sampling and Resolving Power}



A 2.0-pixel sampling is required for the YS, where the sampling is obtained from Full Width at Half Maximum (FWHM) of the Gaussian fit to the collapsed fibre spread function (FSF). This sampling has been shown vary along each diffraction order, with lower sampling at shorter wavelengths of a given order. While the previous iteration of the YS design, V35, was found to be undersampled, V36 has significantly improved on this through modifications to the optical design. The V36 sampling is now over 2 pixels on average. However, it should be noted that these values refer to the ideal, as-designed optical system and thus do not account for manufacturing and alignment tolerances. 

\begin{figure}[htbp]
    \centering
   \includegraphics[height=7.5cm]{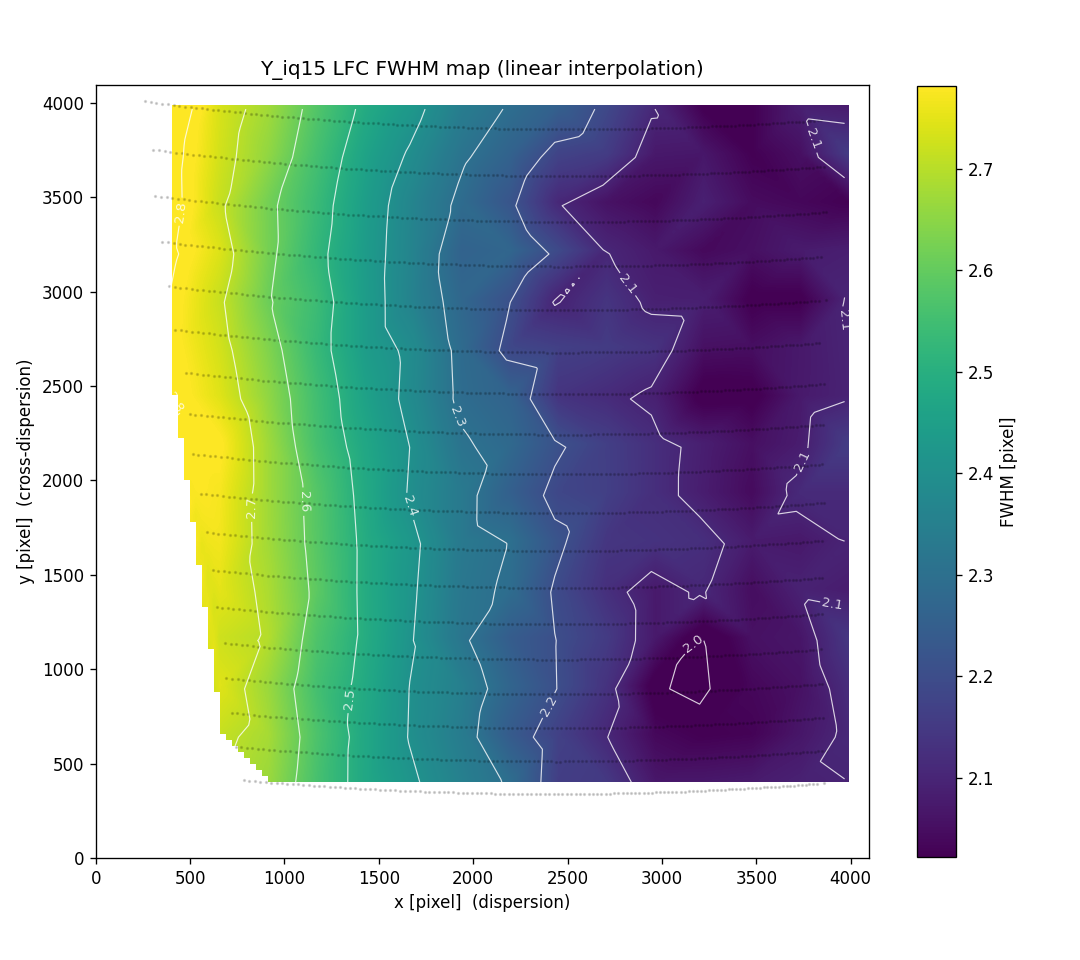}
   \hfill
   \includegraphics[height=7.5cm]{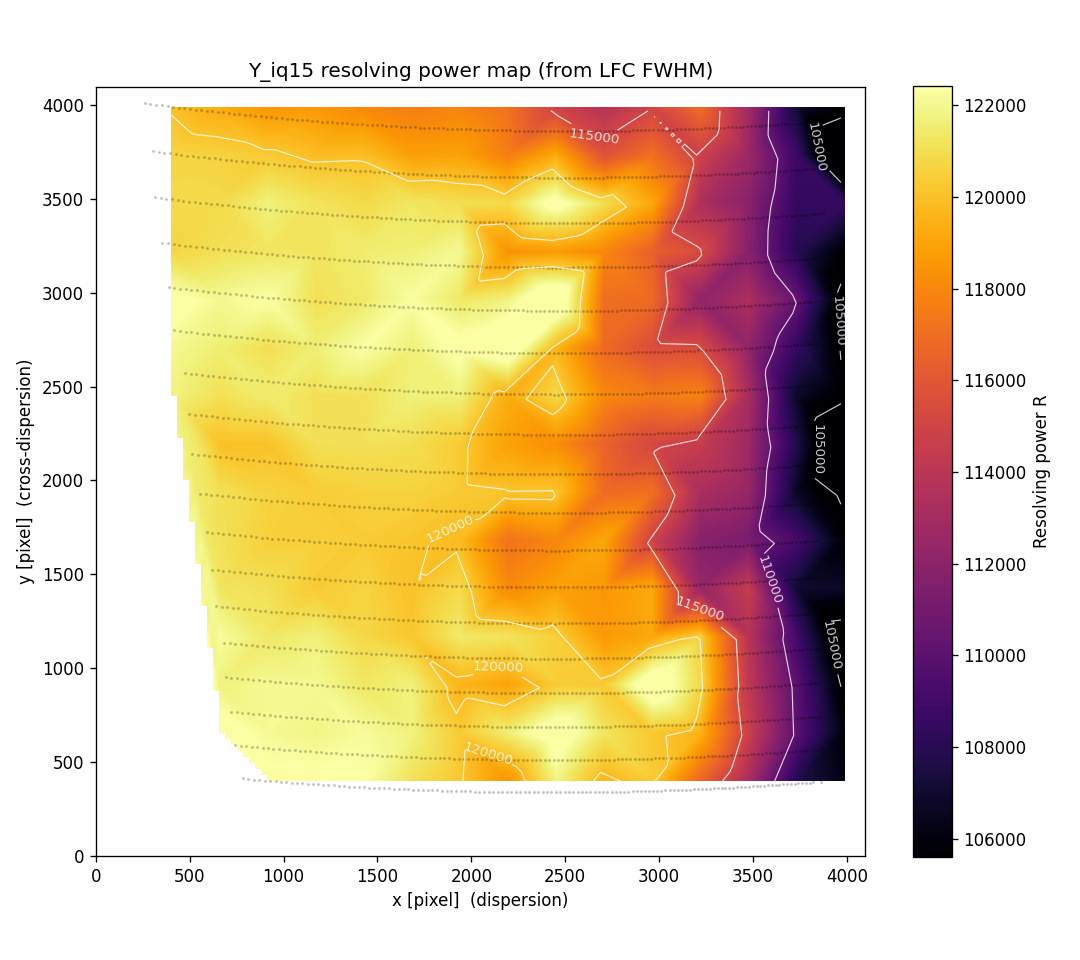}
   \caption[sampling and R for Y] 
   { \label{fig:sampling_r_y} 
Interpolated maps of Sampling (left) and Resolving Power (right) measured in the E2E final product, for Y-band assuming a 4096 × 4096 pixel detector.}
\end{figure}

\begin{figure}[htbp]
    \centering
   \includegraphics[height=7.5cm]{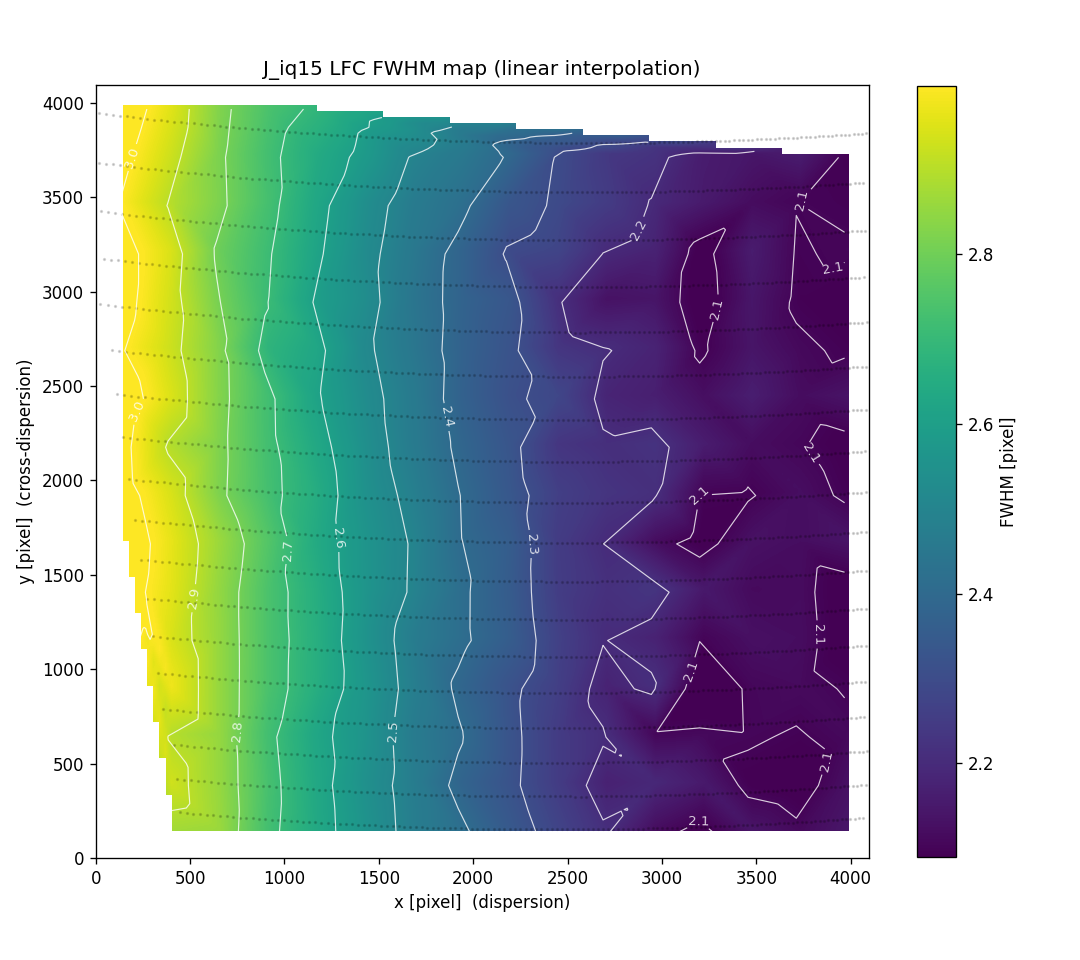}
   \hfill
   \includegraphics[height=7.5cm]{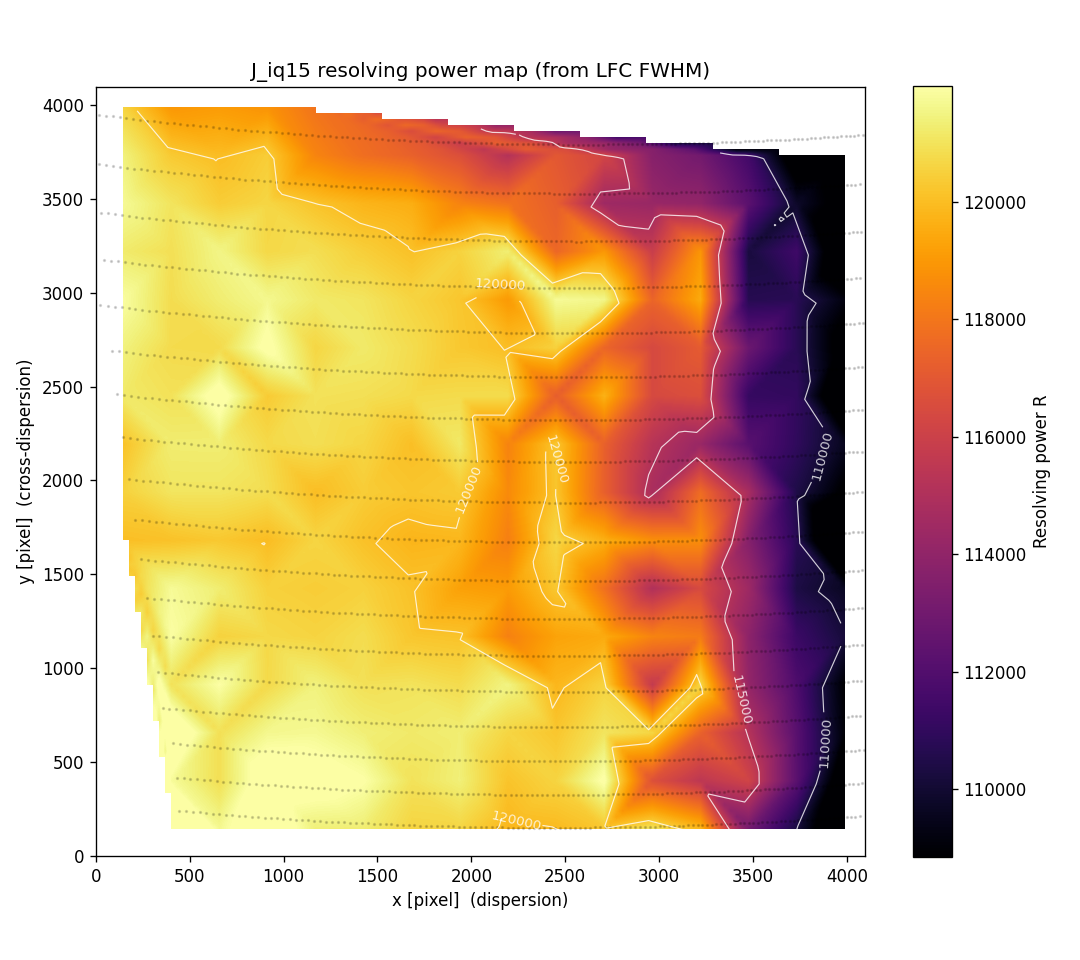}
   \caption[sampling and R for J] 
   { \label{fig:sampling_r_j} 
Interpolated maps of Sampling (left) and Resolving Power (right) measured in the E2E final product, for J-band assuming a 4096 × 4096 pixel detector.}
\end{figure}

\begin{figure}[htbp]
    \centering
   \includegraphics[height=7.5cm]{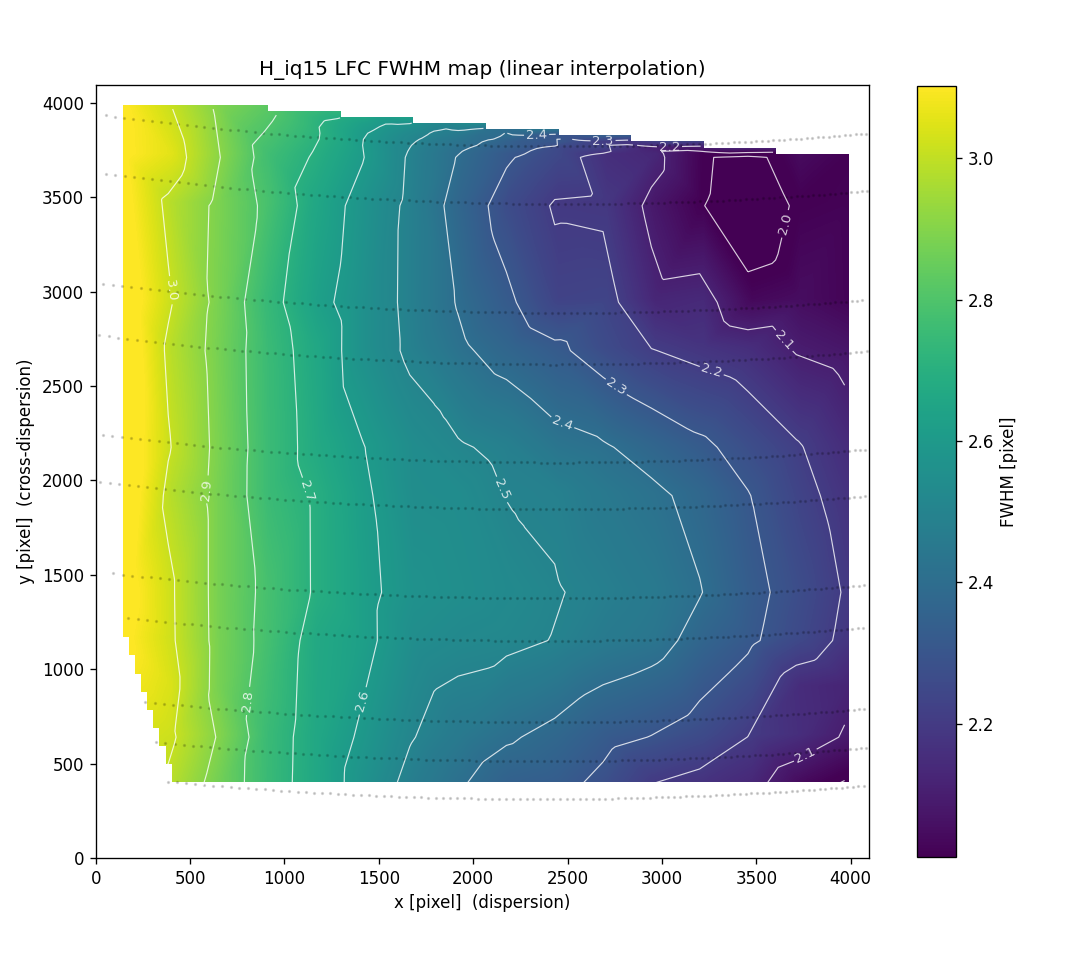}
   \hfill
   \includegraphics[height=7.5cm]{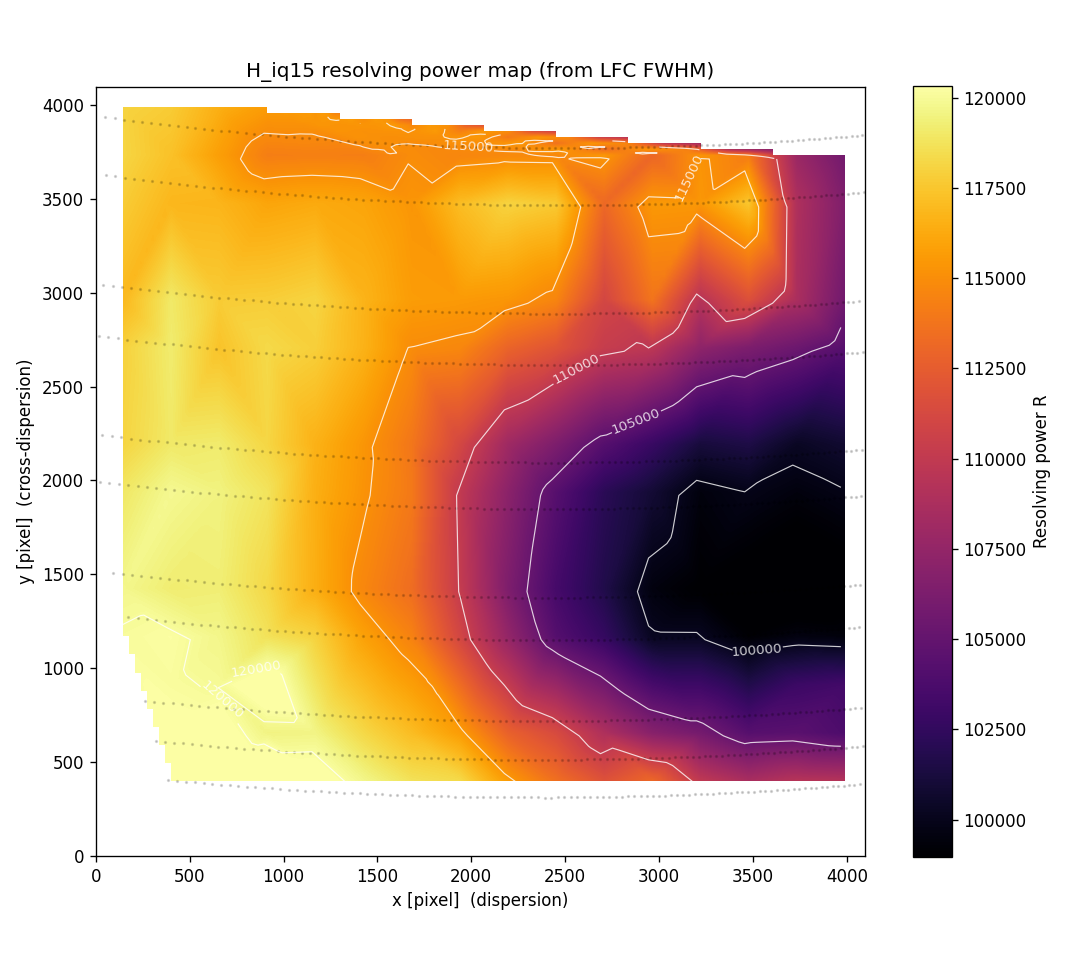}
   \caption[sampling and R for H] 
   { \label{fig:sampling_r_h} 
Interpolated maps of Sampling (left) and Resolving Power (right) measured in the E2E final product, for H-band assuming a 4096 × 4096 pixel detector.}
\end{figure}

Manufacturing tolerances will result in further optical aberrations, degrading the image on the detector and, in turn, resulting in an increased FWHM. Similarly, grating-to-grating alignment tolerances within the mosaic will lead to blurring of the point spread function and an increase in the FWHM. This effect has been simulated, to ensure that the current design would not be intrinsically oversampled, leading to a decrease in resolving power. For this, an estimated 15 \% degradation margin has been applied to represent the as-built spectrometer, based primarily on prior experience with other cryogenic instruments. Fig. \ref{fig:sampling_r_y}, \ref{fig:sampling_r_j} and \ref{fig:sampling_r_h} show the evaluated FWHM and resolving power at the detector from the end-to-end model and using the two-dimensional spectra simulator PyEchelle \cite{pyechelle}. The FWHM now falls between 2.0 and 3.0 pixels at the detector across all three wavebands. Additionally, the inter-order gap exceeds six pixels and the resolving power remains between 100,000 and 123,000 at all wavelengths.

\newpage

\section{Conclusion}

The ANDES YJH spectrograph will operate at the astronomical Y, J, and H bands from 0.95 to
1.8 $\mu$m, providing high-resolution spectra at R = 100,000 over the entire waveband. The YS is designed to support two modes of operation as baseline: a seeing-limited (SL) mode and a diffraction-limited, integral field unit (IFU) mode. A cold slit selector will be employed to select the active slit and block thermal emission from the inactive slit, and a large R4 echelle grating acts as the main disperser of the spectrograph. Due to the capacity of current manufacturing and alignment processes, the grating cannot be fabricated as a single section and must instead be constructed as a grating mosaic. The exact configuration of the mosaic is dependent on the supplier and the grating type selected. The echelle grating of the YJH Spectrograph has been, and remains, the highest risk component of the ANDES YS throughout the project thus far. For this reason, the YJH team has prioritised the de-risking of the R4 echelle grating throughout Phase B and will continue to prioritise it throughout the project.

The current V36 YJH spectrograph design attains an average sampling of 2 pixels. With the consideration of manufacturing and alignment tolerances, which would result in the degrading and blurring the image on the detector, the FWHM of the fibre spread function is expected to increase. Implementing an estimated 15 \% degradation in the end-to-end model of the full instrument, this results in a FWHM between 2.0 and 3.0, with a resolving power between 100,000 and 123,000 at the detector across all three wavebands. These preliminary results will be confirmed in the next phases of the project, with a further matured optical model, full tolerance analysis, and Manufacturing, Assembly, Integration and Test (MAIT) measurements. Any proposed design modifications that may have implications on sampling, image quality, and grating feasibility shall be carefully assessed.


\acknowledgments 

The authors would like to thank Audrey A. Lanotte, Bruno Chazelas, Ulrike Lemke, Daniel Sablowski and Michael Weber for very useful discussions about optical design and modelling. We would also like to thank the ANDES consortium and ESO for their continued support with this project. Lastly, we would like to gratefully acknowledge that this work has been supported by the Science and Technology Facilities Council (STFC).

\newpage
\bibliography{report} 
\bibliographystyle{spiebib} 

\end{document}